\documentclass[12pt,reqno]{amsart}
\usepackage{amssymb}
\usepackage{epsf,epsfig,amsfonts}
\usepackage{mdwlist}
\usepackage{amsfonts,epsfig}
\usepackage[usenames,dvips]{color}

\usepackage[all]{xy}

\usepackage[latin1]{inputenc}

\usepackage{graphicx}

\numberwithin{equation}{section}

\DeclareFontFamily{U}{rsf}{\skewchar\font'177}%
\DeclareFontShape{U}{rsf}{m}{n}{<-6>rsfs5<6-8>rsfs7<8->rsfs10}{}%
\DeclareFontShape{U}{rsf}{b}{n}{<-6>rsfs5<6-8>rsfs7<8->rsfs10}{}%
\DeclareMathAlphabet\RSFS{U}{rsf}{m}{n}
\SetMathAlphabet\RSFS{bold}{U}{rsf}{b}{n}
\newtheorem{thm}{Theorem}[section]
\newtheorem{lem}{Lemma}[section]
\newtheorem{prop}{Proposition}[section]
\newtheorem{corol}{Corollary}[section]
\newtheorem{rem}{Remark}[section]
\newtheorem*{ProbA}{Problem A}
\newtheorem*{ProbB}{Problem B}
\newtheorem{prob}{Problem}
\newcommand{\gG}{g}

\def\sf#1{{\mathsf{#1}}} 

\def\slsf#1{{\slshape \sffamily #1\/}}

\def\smpmatr#1{{\left(\begin{smallmatrix} #1 \end{smallmatrix}\right)}}

\def\state#1. {\noindent{\bf{#1.} }}

\def\cc{{\mathbb C}} 
\def\rr{{\mathbb R}}

\def\cald{{\mathcal{D}}}
\def\cale{{\mathcal{E}}}
  
\def\calf{{\mathcal{F}}}

\def\call{{\mathcal{L}}}

\def\half{{\frac12}}

\def\vph{^{\mathstrut}}

\let\wt=\widetilde

\def\bs{\backslash}
\def\const{{\sf{const}}}

\def\cplxi{{\mskip2mu\sf{i}\mskip2mu}}

\def\sin{{\sf{sin}}}
\def\cos{{\sf{cos}}}
\def\sinh{{\sf{sinh}}}
\def\cosh{{\sf{cosh}}}
\def\exp{{\sf{exp}}}

\def\log{{\sf{log}}}
\def\lim{{\sf{lim}}}

\def\inv{^{-1}}
\let\ti=\tilde

\def\mbfv{{\boldsymbol{v}}}

\def\leq{\leqslant}
\def\geq{\geqslant}

\def\sli{{\sl i)\ \;}}\def\slip{{\sl i)}}
       \def\sliip{{\sl i$_{\!}$i)}}\def\slii{{\sliip\ \;}}
\def\sliiip{{\sl i$_{\!}$i$_{\!}$i)}}
       \def\sliii{{\sliiip\ \;}}

\newcommand{\weg}[1]{}

{\catcode`\@=11
\gdef\n@te#1#2{\leavevmode\vadjust{%
 {\setbox\z@\hbox to\z@{\strut#1}%
  \setbox\z@\hbox{\raise\dp\strutbox\box\z@}\ht\z@=\z@\dp\z@=\z@%
  #2\box\z@}}}
\gdef\leftnote#1{\n@te{\hss#1\quad}{}}
\gdef\rightnote#1{\n@te{\quad\kern-\leftskip#1\hss}{\moveright\hsize}}
\gdef\?{\FN@\qumark}
\gdef\qumark{\ifx\next"\DN@"##1"{\leftnote{\rm##1}}\else
 \DN@{\leftnote{\rm??}}\fi{\rm??}\next@}}

\definecolor{DarkGreen}{rgb}{0,0.5,0} 
\definecolor{DarkBlue}{rgb}{0,0,0.6}

\definecolor{MyMagenta}{rgb}{0.6,0.0,0.5}

\begin{document}

\baselineskip =14.7pt plus 2pt
\leftmargini=27pt 
\leftmarginii=21pt 
\leftmarginiii=19pt 
\leftmarginiv=19pt  

\title[Two-dimensional metrics with linear and cubic integrals]%
{Two-dimensional  superintegrable metrics with    \\[4pt] one linear and  one  cubic integral.}\author[V.~S.~Matveev]{Vladimir  S.~Matveev}
\address{Mathematisches Institut\\ 
 Friedrich-Schiller-Universität Jena\\ 
 07737  Jena } 
\email{vladimir.matveev@uni-jena.de}
\author[V.~V.~Shevchishin]{Vsevolod V.~Shevchishin}
\address{Iniversität Hamburg\\
Department Mathematik\\
Bundesstr. 55\\
 20146 Hamburg\\
Germany
} \email{shevchishin@googlemail.com}\thanks{  
the first author was partially supported by  DFG (SPP 1154 and GK 1523)}


\begin{abstract}
 We describe all { local} Riemannian metrics on surfaces whose geodesic
 flows are superintegrable with one integral linear in momenta and one
 integral cubic in momenta.

{ We also show that some of these metrics can be extended to $S^2$. This
gives us  new examples of  Hamiltonian systems on the sphere
with integrals of degree three  in momenta, and the first examples of superintegrable  metrics of nonconstant curvature on a closed surface.}
\end{abstract}
\maketitle

\section{Introduction}
\subsection{Definitions and statement of the problem} \label{intr1} Let $M^2$ be
a surface (i.e., 2-dimensional real manifold) equipped with a Riemannian
metric $g=(g_{ij})$. The \slsf{geodesic flow} of the metric $\gG$ is the
Hamiltonian system on the cotangent bundle  $T^*M^2$ with the Hamiltonian
$H:=\tfrac{1}{2}g^{ij}p_ip_j$, where $(x,y)=(x_1, x_2)$ is a local coordinate
system on $M^2$, and $(p_x, p_y)=(p_1,p_2)$ are the correspondent
\slsf{momenta}, i.e., the dual coordinates on $T^*M^2$.

We say that a function $F:T^*M^2\to \mathbb{R}$ is an \slsf{integral} of the
geodesic flow of $g$, if $ \{F, H\}=0$, where $\{ \ , \ \}$ is the canonical
Poisson bracket on $T^*M^2$.  We say that the integral is \slsf{polynomial in
 momenta of degree $d$}, if in every local coordinate system $(x,y,p_x,p_y)$
it has the form
\begin{equation} \label{eqn:int}
\textstyle
  F(x,y,p_x,p_y)= \sum_{i=0}^da_i(x,y) p_x^{d-i}p_y^i, 
\end{equation}

For example, the  Hamiltonian $H$ itself is an integral quadratic in momenta. 
Integrals polynomial in momenta of degree 1 (3, resp.) will be called \slsf{linear} (resp. \slsf{cubic}) integrals.

The first main result of the present paper is  a complete solution of  the following problem: 

\begin{ProbA}  {\it Describe locally all two-dimensional
  Riemannian metrics admitting one integral $L$ linear in momenta and one
  integral $F$ cubic in momenta such that $L, F,$ and $H$ are functionally
  independent.} \end{ProbA}

Recall that functions $L, F, H$ are \slsf{functionally independent} if there
exists a point on $T^*M$ such that at this point the differentials $dL$, $dF$,
$dH$ are linearly independent. For integrals polynomial in momenta, linear
independence of the differentials of the integrals at one point implies linear
independence of the differentials of the integrals at
 every point of a certain everywhere dense open subset
(assuming the manifold is connected).

Recall that two-dimensional metrics whose geodesic flows admit three
functionally independent integrals of a certain special form (in most cases the integrals are assumed to be polynomial in momenta of certain fixed degrees) 
 are called
\slsf{superintegrable}. Superintegrable metrics (and Hamiltonian systems in
general) is nowadays a hot topic in mathematical physics and differential
geometry, due to various applications and interesting mathematical structure
lying behind. We {suggest}
\cite{K-K-M-W,K-K-M,Te-Wi,Da-Yp} for a discussion of superintegrable systems
from the viewpoint of mathematical physics, and \cite{Kr,Br-Ma-Ma} from the
viewpoint of differential geometry.

If the metric is superintegrable with two linear integrals, it has constant
curvature. The metrics that are superintegrable with two quadratic integrals
(in addition to the energy integral), or one linear and one quadratic, were
described (locally, in a neighbourhood of almost every point) in the classical
work of Koenigs \cite{Koe}.

The next case should be ``linear integral + cubic integral'', but the only
result we found in this direction is due to Rañada
\cite{Ra}, Gravel \cite{Gr}, and Marquette and Winternitz \cite{Mar-Wi} and
concerns the Hamiltonian systems such that the Hamiltonian $H$ is the sum of
the standard kinetic energy $K= \tfrac{1}{2} (p_x^2 + p_y^2) $ and a potential
energy $V(x,y)\neq \const $. They assumed the existence of (functionally
independent) linear and cubic integrals and proved that for such systems the
cubic integral is actually the product of the linear integral and of an
integral quadratic in momenta, i.e., such systems can be obtained via the
Maupertuis' transformation from the superintegrable systems constructed by
Koenigs \cite{Koe}.  In particular, all known examples of metrics satisfying
assumptions in Problem A   above were in a certain sense trivial: the metric has a
constant curvature, or the metric is superintegrable with one quadratic and one
linear integral, and every cubic integral is a product of the integral linear
in momenta and  an integral quadratic in momenta.

\subsection{Main result: local normal forms of metrics admitting one
 linear and one cubic integral}

\begin{thm} \label{main.th} Let $g$ be a Riemannian metric on the
 2-dimensional connected manifold $M^2$. 

 Suppose the geodesic flow of $g$ admits a linear integral $L$ and a cubic
 integral $F$ such that $L$, $F$ and the Hamiltonian $H$ are functionally
 independent. Then, locally near every point $p$ such that
  $L_{|T^*_pM^2} \not\equiv 0$ there exist coordinates $(x,y)$ and a real function
  $h(x)$ such that the metric $g$ has the form $g=\tfrac{1}{h_x^2}(dx^2+dy^2)$
  where $h_x$ is the $x$-derivative, and the function satisfies one of the
  following \slsf{Principal} (ordinary differential)  \slsf{equations}:
\begin{equation}\label{Eqh1all}
\begin{split} 
\textstyle 
\ \ \ \,\llap{\text{(i)\ \,}}
 h_x{\cdot}\bigr(A_0{\cdot} h_x^2 +\mu^2{\cdot}A_0{\cdot}h(x)^2 -A_1{\cdot}h(x)+A_2 \bigr)
&\textstyle 
\;- \bigl( A_3{\cdot}\frac{\sin(\mu{\cdot}x)}{\mu}\  +A_4{\cdot}\cos(\mu{\cdot}x) \bigr) \;\,=0\ \ 
\\
\llap{\text{(ii)\ \,}}
 h_x{\cdot}\bigr(A_0{\cdot}h_x^2 -\mu^2{\cdot}A_0{\cdot}h(x)^2 -A_1{\cdot}h(x)+A_2 \bigr)
&\textstyle 
\;- \bigl( A_3{\cdot}\frac{\sinh(\mu{\cdot}x)}{\mu} +A_4{\cdot}\cosh(\mu{\cdot}x) \bigr) =0
\\
\llap{\text{(iii)\ \,}}
 h_x{\cdot}\bigr( A_0{\cdot}h_x^2 \hphantom{\hbox{$-\mu^2{\cdot}A_0{\cdot}h(x)^2$\ \,}} 
-A_1{\cdot}h(x)+A_2 \bigr)
&\;- \bigl( A_3{\cdot}x
\hphantom{\hbox{$\sinh(\ \,$}} +A_4 \bigr) =0
\end{split} 
\end{equation}
with $\mu>0$ in the  first two cases.

In all three cases the metric $g=\tfrac{1}{h_x^2}(dx^2+dy^2)$ is
superintegrable with one linear integral $L=p_y$ and one cubic
 integral.  In the case (i) a cubic integral $F=F(x,y;p_x,p_y)$ can be given
by
\[\textstyle
F =( C_+{\cdot}e^{\mu y}+ C_-{\cdot}e^{-\mu y}){\cdot}
\big(a_0(x){\cdot}p_x^3+a_2(x){\cdot}p_x\vph p_y^2\big) +
( C_+{\cdot}e^{\mu y}- C_-{\cdot}e^{-\mu y}){\cdot}\big(a_1(x){\cdot}p_x^2p_y\vph  +a_3(x){\cdot}p_y^3\big) 
\]
where  $C_+,C_-$ are arbitrary   constants   and  $a_i(x)$ are functions  given by 
\begin{equation}\label{a0-a3new}
\textstyle 
\begin{split}
a_0(x) &\textstyle 
= A_0{\cdot}h_x^3\\
a_1(x) &\textstyle 
= (-\mu{\cdot}A_0{\cdot}h(x)+\frac{A_1}{2\mu}){\cdot}h_x^2\\
a_2(x) &\textstyle 
=  \,\tfrac{1}{2}\,{\cdot}(3A_0{\cdot}h_x^2+ \mu^2{\cdot}A_0{\cdot}h(x)^2  
-A_1{\cdot}h(x) + A_2){\cdot}h_x  \\
a_3(x) &\textstyle 
= \tfrac{1}{2\mu}{\cdot}(3A_0{\cdot}h_x^2
+\mu^2{\cdot}A_0{\cdot}h(x)^2-A_1{\cdot}h(x)+A_2){\cdot}h_{xx}.
\end{split}
\end{equation} 

In the case (ii) { a  cubic integral can be given by}
\[
F= C_e{\cdot}\cos(\mu{\cdot}y+\phi_0){\cdot}\big(a_0(x){\cdot}p_x^3+a_2(x){\cdot}p_x\vph p_y^2\big) +
C_e{\cdot}\sin(\mu{\cdot}y+\phi_0){\cdot}\big(a_1(x){\cdot}p_x^2p_y\vph  +a_3(x){\cdot}p_y^3\big) 
\]
where $C_e,\phi_0$ are constants (``amplitude and phase'') and $a_i(x)$ are
functions given by
\begin{equation}\label{a0-a3-ell}
\textstyle 
\begin{split}
a_0(x) &\textstyle  = A_0{\cdot}h_x^3\\
a_1(x) &\textstyle  = (\mu{\cdot}A_0{\cdot}h(x)+\frac{A_1}{2\mu}){\cdot}h_x^2\\
a_2(x) &\textstyle 
=  \,\tfrac{1}{2}\,{\cdot}(3A_0{\cdot}h_x^2- \mu^2{\cdot}A_0{\cdot}h(x)^2  
-A_1{\cdot}h(x) + A_2){\cdot}h_x  \\
a_3(x) &\textstyle 
= \frac{1}{2\mu}{\cdot}(3A_0{\cdot}h_x^2
- \mu^2{\cdot}A_0{\cdot}h(x)^2-A_1{\cdot}h(x)+A_2){\cdot}h_{xx}.
\end{split}
\end{equation}
In the case (iii) { a  cubic integral can be given by}
\begin{equation}\label{Fmu0}
\begin{split}
F=\;&
\textstyle 
 C_1{\cdot}\Big( a_0(x){\cdot}p_x^3  + a_2(x){\cdot}p_x^{\,}p_y^2
\; +\frac{y}{2}{\cdot}\big(A_1{\cdot}h_x^2{\cdot}p_x^2p_y^{\,}   
+ (A_1{\cdot}h_x^2+A_3){\cdot} p_y^3\big)  \!\Big) 
\\
+\;&
\textstyle 
 C_2{\cdot} \Big( y{\cdot}a_0(x){\cdot}p_x^3  
+ a_1(x){\cdot}p_x^2p_y^{\,}  
+ y{\cdot}a_2(x){\cdot}p_x^{\,}p_y^2 
+ a_3(x){\cdot}p_y^3\\
& \textstyle
\qquad\quad +\frac{y^2}{4}{\cdot}\big(A_1{\cdot}h_x^2{\cdot}p_x^2p_y^{\,}   
+ (A_1{\cdot}h_x^2+A_3){\cdot} p_y^3\big)\Big)
\end{split}
\end{equation}
where  $C_1,C_2$ are  constants and $a_i(x)$ are functions  given by 
\begin{equation}\label{a0-a3mu0}
\textstyle 
\begin{split}
a_0(x) &\textstyle  = \ \;\, A_0{\cdot}h_x^3\\
a_1(x) &\textstyle  = - A_0{\cdot}h_x^2{\cdot}h(x)\\
a_2(x) &\textstyle 
=  \half{\cdot}(3{\cdot}h_x^2{\cdot}A_0  
-A_1{\cdot}h(x) + A_2){\cdot}h_x  \\
a_3(x) &\textstyle  
= -\frac14{\cdot}(4A_0{\cdot}h_x^2{\cdot}h(x) +A_3{\cdot}x^2 +2A_4{\cdot}x ) . 
\end{split}
\end{equation}

\smallskip %
Moreover, in the case when the metric $g$ has non-constant curvature every 
cubic integral is a linear combination $F+C_{L3}{\cdot}L^3+C_{LH} {\cdot}L{\cdot}H$ where
$F$ is given by the above formula (according to the case (i)--(iii)) and
$C_{L3},C_{LH}$ are  constants. In particular, in the non-constant
curvature case the space of cubic integrals of our metrics  has dimension $4$.
\end{thm}

\begin{rem}\label{rem-uniq} \bf Uniqueness of the equation.
 \rm We show in Theorem \ref{thm-uniq} that in the case when the curvature 
 of our  metric $g=\tfrac{1}{h_x^2}(dx^2+dy^2)$ is non-constant the equation
 \eqref{Eqh1all} on  the function $h(x)$ is \emph{unique} up to a  constant
 factor.  On the other hand, in Theorem \ref{R=const} we describe possible
 equations of the form \eqref{Eqh1all} for whose  the metric
 $g=\tfrac{1}{h_x^2}(dx^2+dy^2)$ has constant curvature.

Thus, Theorems \ref{main.th}, \ref{R=const}, and \ref{thm-uniq} give a
complete answer to the Problem A   above.
\end{rem}

\begin{rem}\label{rem-Darboux} \bf Known special case:
 Darboux-superintegrable metrics. %
 \rm We call a metric $g$ on $M^2$ \slsf{Darboux-superintegrable}, if it has
 non-constant curvature and the geodesic flow of the metric admits at least
 four linear independent integrals quadratic in momenta.  For example, such is
 the metric $(x^2+y^2+1){\cdot}(dx^2+dy^2)$ on $\rr^2$ (see for example
 \cite[§4]{Ma1}).

 Darboux-superintegrable metrics are well understood.  Locally, they were
 described already by Koenigs \cite{Koe}. In particular, Koenigs has shown
 that every Darboux-superintegrable metric admits a linear integral.  Then, it
 also admits  cubic integrals, namely  the products of the linear integral and the
 quadratic integrals.   Therefore, our Theorem \ref{main.th}
 applies. In particular, in  the appropriate local coordinates $(x,y)$ the metric
 has the form $\frac{1}{h_x^2}(dx^2+dy^2)$ such that $h(x)$ is a solution of
 the Principle equation \eqref{Eqh1all}, (i)--(iii).

 The formulas above show that if the
 coefficient $A_0$   vanishes, then a  generic 
  cubic integral  $F+C_{L3}{\cdot}L^3+C_{LH} {\cdot}L{\cdot}H$ is the    product  of  the 
  linear  integral and a function quadratic in momenta 
   which must automatically be an integral. Then, the metrics corresponding to $A_0=0$ are  Darboux-superintegrable or 
are of constant curvature.       The uniqueness of the Principle equation (see
 the previous remark) shows that the converse assertion is also true. Thus we
 obtain the following characterisation: {\it Under the hypotheses of Theorem
  \ref{main.th} the metric $g=\frac{1}{h_x^2}(dx^2+dy^2)$ { of
   non-constant curvature} is Darboux-superintegrable if and only if the
  parameter $A_0$  vanishes.}
\end{rem}

\begin{rem} {\rm  With the help of a computer algebra software, for example with  Maple\textsuperscript{®}, it is easy to check that the functions $F$ from Theorem \ref{main.th} are indeed integrals: the condition $\{H, F\}=0$ is equivalent to 5 ODEs of at most 3rd order on the function $h$; these ODEs are  identically fulfilled for the function $h$  satisfying the corresponding equation \eqref{Eqh1all}, since they are algebraic corollaries of the corresponding  equation and its first two  derivatives. We will of  course explain how we constructed the integrals (in Section \ref{S2}) since we also need to show  that we constructed all such metrics. Moreover, the idea of the construction   will be used in  the proof of other statements of Theorem \ref{main.th}, in particular in the proof that the dimension of the space of the integrals is 4. Moreover, we believe that the idea of our construction could also be used for constructing 
 higher order superintegrable cases, see Problem \ref{p1} in the conclusion.}\end{rem}

\subsection{Second main result: Examples of 
 metrics on the  $2$-sphere admitting  linear and cubic integrals}\label{glob-intro}

 The
problem of finding and describing \emph{global} integrable Hamiltonian
systems, i.e., those whose configuration  space is a compact manifold, is one of 
the central topics  in the classical mechanics.  
 The version of this problem in  our
context is as follows:

\begin{ProbB} %
 \label{prob-glob}
 {\it Understand what  Riemannian metrics  
 on the 2-sphere  $S^2$ admit  one integral $L$ linear in momenta and one   integral $F$ cubic in momenta such that $L, F,$ and the Hamiltonian $H$ are functionally
  independent.} \end{ProbB}
  
  Note that other oriented closed surfaces can not admit superintegrable metrics. Indeed, if the metric is superintegrable, all geodesics are closed which is possible on the sphere and on $\mathbb{R}P^2$ only.  
  
It is known that   the   metric of constant curvature of the sphere do admit (linearly independent)  linear and cubic integrals.   So the nontrivial part of the Problem B is whether there are other metrics on the 2-sphere  admitting 
an integral $L$ linear in momenta and an    integral $F$ cubic in momenta such that $L, F,$ and $H$ are functionally independent.

For the integrals of lower degrees, the answer in negative. Indeed, the existence of two functionally  independent linear integrals implies, even locally, that the metric has constant curvature. By Kiyohara \cite{Ki},  the existence of three functionally 
independent  quadratic integral (energy integral + two additional integrals) on the 2-sphere implies that the metric is of constant curvature. From this result, it also follows that the existence of (functionally independent) linear  and two quadratic   integrals  implies that  the metric of the 2-sphere 
 has constant  curvature.  Because of these results (and absence of  examples of  superintegrable systems with higher degree integrals), it was  generally believed that no polynomially superintegrable metric  exists on a closed surface of nonconstant curvature.

 In the present paper, we  construct the first examples of  smooth (even analytic) metrics  of nonconstant curvature 
 on the 2-sphere    whose geodesic flows admit  integrals $L$ linear in momenta and    integrals $F$ cubic in momenta such that $L, F,$ and $H$ are functionally independent. 
 The construction is in Section
\ref{globalism}. We show that for certain values of parameters the metrics we constructed  in Theorem \ref{main.th} and the integrals of these metrics can be  smoothly extended  to the sphere. More precisely (in the notation of Theorem \ref{main.th}), 
 if the function $h(x)$ fulfills the equation
\eqref{Eqh1all} (ii) and the condition $h'(x_0)>0$ at some point $x_0$ whereas
the real parameters $\mu>0,A_0,\ldots,A_4$ satisfy inequalities $A_0>0$,
$\mu{\cdot}A_4>|A_3|$ then the metric $g=\frac{1}{h_x^2}(dx^2+dy^2)$ smoothly
extends to the sphere $S^2$ together with the linear integral $L=p_y$ and the
cubic integral $F$ given by \eqref{a0-a3-ell}. 

We  conject that {\it these are all examples of metrics on the sphere superintegrable with one linear and one cubic integral}. 

Our  examples  are also  interesting from other points of view. Indeed, every  metric from these examples   admit and  integral cubic in momenta that  not the product of a linear and a quadratic integral. The problem of constructing such metrics   is very classical and, 
 in a certain extend,   was stated  by Jacobi,  Darboux, Cauchy, Whittaker, 
see also  \cite{Bo-Ko-Fo,Bo-Fo}.     There are only very few examples of such metrics on closed surfaces:  constant curvature metrics, 
metrics constructed via Maupertuis' transformation from the  Goryachev-Chaplygin case of rigid body motion and their generalizations due to Goryachev \cite{Go},    metrics constructed by Selivanova \cite{Se},  by Kiyohara \cite{Ki2}, and by Dullin at al \cite{Du-Ma} (see also \cite{Du-Ma-To} and \cite{Ma-Sh}). Note that the analogous question for the quadratic integrals is completely solved, see \cite{Kol,Ki,Bo-Ma-Fo,Ma}. 

Moreover, all 
geodesics of  the metrics we constructed  are closed  (since it is always the case for superintegrable metrics), so the examples are also examples of the so-called Zoll surfaces.

\subsection{Additional  result: Special case of Kruglikov's  ``big gap'' conjecture.} \label{5} 

In \cite[§12]{Kr}, Kruglikov has shown that the dimension of the space of
cubic integrals (of the geodesic flow of a 2D-metric) is at most 10; the
dimension 10 is achieved only by the metrics of constant curvature. He also
has shown that the second largest dimension is at most 7 (see \cite[Theorem
8]{Kr}), and conjectured that the gap between the largest and the second
largest possibilities for the dimension of the space of cubic integrals is
even bigger: he writes that {\it it seems that the next realized dimension
 after 10 is 4}.
 
We will prove this conjecture (see Theorem \ref{thm-4}) under the additional
assumption that the metric admits a Killing vector field. Note that this
assumption does not look too artificial, since it is expected
that metrics with many polynomial integrals admit Killing vector fields. For
example, by the classical result of Koenigs { mentioned} above, metrics
admitting four  (= the second largest dimension) 
linearly independent integrals that are quadratic in momenta
admit Killing vector fields.

More precisely, we will prove that, if a 2D metric is superintegrable with one
linear and one cubic integral ($L$ and $F$) and has non-constant curvature,
then, locally, the space of cubic integrals is precisely 4-dimensional. In particular,  
  in
addition to the integrals $L^3, F, L\cdot H$ we always construct one more cubic
integral $F_2$ that is linearly independent of $L^3, F,$ and $ L\cdot H$.

\section{ Principle equation and overview of the proof of Theorem
 \ref{main.th} } \label{S2}
 \subsection{  How we found  the metrics: scheme of the proof of Theorem  \ref{main.th}.} \label{1.2} 

 It is well-known (see for example \cite[§592]{Da}, or \cite{Bo-Ma-Fo}) that
 every pair $(g,L)$, where $g$ is a Riemannian metric, and $L$ is an integral
 linear in momenta, is given { in appropriate coordinates} in a neighbourhood
 of every point such that $L\not\equiv0$ by the formulas
\begin{equation} \label{lin.int}
g= \lambda(x) (dx^2+ dy^2) \,  \textrm{ and} \   \   L= p_y.
\end{equation}

The natural ``naive'' method to solve the Problem A  
  would be to write the
condition $\{H,F\}=0$, where $H = \tfrac{p_x^2 + p_y^2}{2\lambda(x)}$ and %
$F:= a_0(x,y)p_x^3 +a_1(x,y)p_x^2p_y + a_2(x,y)p_xp_y^2 + a_3(x,y)p_y^3$, as
the systems of PDE on the unknown function $\lambda$ of one variable and unknown
functions $a_i$ of two variables, and to try to solve it.  Unfortunately, by
this method we obtain a system of 5 nonlinear PDE on 5 unknown functions
$\lambda,a_0,a_1,a_2,a_3$, which is completely intractable\footnote{This
 ``naive'' approach to this problem was tried without success by many experts
 in superintegrable systems (private communications by Marquette,
 Rañada, Winternitz).}.

In order to solve the problem, we used a trick that allowed us to reduce the
problem to solving systems of ODE (instead of PDE).  A similar trick was
recently used in \cite{Ma2}.
 
The main observation is the following: the Poisson bracket of the linear
integral $L$ and of a cubic integral $F$ is
\begin{itemize}
\item an integral (because of the Jacobi identity), and 
\item is cubic in momenta (because each term in the sum %
 $\{L, F\} = \partial_x F \partial_{ p_x } L + \partial_y F \partial_{ p_y} L- \partial_x L \partial_{ p_x } F - \partial_y L \partial_{
  p_y} F$ is cubic in momenta). 
\end{itemize}
Thus, the mapping $\call: F \mapsto \{L, F\}$ is a linear homomorphism. By \cite{Kr}, the space of cubic integrals is
finite- (at most, 10-)dimensional. Let us now consider the eigenvalues of the
mapping $\call$. Clearly, $0$ is an eigenvalue of $\call$, whose eigenvectors are %
$A_3 \cdot L^3 + A_1\cdot L \cdot H$, where $A_1, A_3 \in \mathbb{R}$.  The following two
cases are possible:

\medskip\noindent %
{\bf Case 1: The mapping $\call$ has an eigenvalue $\mu \neq 0$.}  Then, there exists a
cubic integral $F$ such that $\{L, F\}= \mu \cdot F $. We allow $\mu$ to be a complex
number, and $F$ to be complex valued function, i.e., $F= F_1 + \cplxi F_2$ for
real-valued cubic integrals $F_1$ and $F_2$.
 
In the coordinates such that $(g, L)$ are given by \eqref{lin.int}, we
have 
\[
\{L, F\}= \partial_y F= \partial_y a_0(x,y) \cdot p_x^3 +\partial_y a_1(x,y)\cdot p_x^2p_y + \partial_y
a_2(x,y)\cdot p_xp_y^2 + \partial_y a_3(x,y) \cdot p_y^3,
\]
so that the condition $\{L, F\}=\mu \cdot F$ is equivalent to the system
$ \partial_y
a_i(x,y) = \mu \cdot a_i(x,y)$, $i=0,...,3. $ Then, $a_i(x,y)= \exp(\mu y)\cdot a_i(x)$
for certain (complex valued in the general case) functions $a_i(x)$ of one
variable $x$. Then, all unknown functions in the equation $\{H, F\}=0$ are
functions of the variable $x$ only, i.e., the condition $\{H, F\}=0$ is a system
of ODE (depending on the parameter $\mu$).  Finally, the condition $\{H, F\}=0$ is
equivalent to $5$ ODE on $5$ unknown functions of one variable $x$: four
unknown functions  $a_i(x)$ and $\lambda(x)$. Working with this system of ODE, we
partially integrate it and reduce it to one ODE of the first order
(essentially, the first equation of \eqref{Eqh1all} for $\mu\in \mathbb{R}$ and
the second equation of \eqref{Eqh1all} for $\mu\in \cplxi\cdot \mathbb{R}$. Further in
§\ref{real-sols} we shall show that the assumption that the function $h_x$ is
real implies that $\mu$ is real or pure imaginary).

\begin{rem} \rm In this way, we obtain $\lambda(x)$ which is a priori a complex-valued
  function; for our problem, only real-valued $\lambda$'s are of interest.  We shall 
  see in §\ref{real-sols} that $\lambda$ is real if and only if $\mu$ is real or purely
  imaginary.
\end{rem}

\medskip\noindent %
{\bf Case 2: The mapping $\call$ has only one eigenvalue, namely zero.}  Since
in our setting the space of cubic integrals is at least three-dimensional,
there exists an integral $F$ linear independent of $L^3 $ and $L\cdot H$ such that
$ \{L, F\}= \tfrac{A_3}{2} \cdot L^3 + A_1 \cdot L \cdot H$ for certain constants $A_1,A_3$.
In the coordinates such that $(g,L)$ is given by \eqref{lin.int}, the
condition $ \{L, F\}= \frac{A_3}{2} \cdot L^3 + A_1 \cdot L \cdot H$ reads
\begin{eqnarray*} & & \{p_y, a_0(x,y)p_x^3 +a_1(x,y)p_x^2p_y + a_2(x,y)p_xp_y^2 + a_3(x,y)p_y^3\} \\ 
   &=&   \partial_y a_0(x,y) \cdot p_x^3 +\partial_y  a_1(x,y)\cdot p_x^2p_y + \partial_y  a_2(x,y)\cdot p_xp_y^2 + \partial_y  a_3(x,y) \cdot p_y^3 \\ 
   & =&   \tfrac{A_3}{2}\cdot  p_y^3 + \tfrac{A_1}{2\lambda(x)} \cdot (p_x^2  + p_y^2)\cdot p_y, 
\end{eqnarray*}
and is equivalent to the system $\partial_y a_0(x,y) = 0 $,
$\partial_ya_1(x,y)=\frac{A_1}{2\lambda(x)}$, $\partial_y a_2(x,y) = 0$, $\partial_y a_3(x,y) =
\tfrac{A_3}{2}+ \frac{A_1}{2\lambda(x)}$. Then,
\begin{equation}  \label{a01-mu0}
F= a_0(x){\cdot}p_x^3 +a_1(x){\cdot}p_x^2p_y + a_2(x){\cdot}p_xp_y^2
+ a_3(x){\cdot}p_y^3 + \tfrac{y}{2}\cdot \left(A_3 {\cdot} p_y^3 
+ A_1 {\cdot} p_y {\cdot} \tfrac{p_x^2 +  p_y^2}{\lambda(x)} \right).
\end{equation} 
We again see that all unknown functions in the equation $\{H, F\}=0$ are
functions of the variable $x$ only, i.e., the condition $\{H, F\}=0$ is a system
of 5 ODE (depending on the parameters $A_1, A_3, y_0$) on 5 unknown functions
of one variable $x$: $a_i(x)$ and $\lambda(x)$. Working with this system of ODE, we
partially integrate it and reduce it to one ODE of the first order (Equation (iii)
of \eqref{Eqh1all}), which is in a certain sense a degenerate case of the
corresponding ODE we obtained in Case 1.

\subsection{Case 1 ($\mu \neq 0$) in the proof of Theorem \ref{main.th}}\label{Case1} 
For convenience in further computation, we write the metric $g$  in the form 
\begin{equation}\label{gh}
\textstyle
g=\frac{dx^2+dy^2}{h_x^2}
\end{equation} 
for some function $h=h(x)$, where $h_x=\frac{dh(x)}{dx}$. Then
$H=\frac{h_x^2}{2}{\cdot}(p_x^2+p_y^2)$ and the linear integral is $L:= p_y$.

We assume (see §\ref{1.2}) that there exists a complex-valued cubic integral
of the form
\[
F= \exp(\mu {\cdot} y){\cdot} a_0(x) {\cdot} p_x^3 +\exp(\mu {\cdot} y) {\cdot} a_1(x){\cdot} p_x^2p_y\vph +
\exp(\mu {\cdot} y) {\cdot} a_2(x){\cdot} p_x\vph p_y^2 + \exp(\mu {\cdot} y) {\cdot} a_3(x) {\cdot} p_y^3,
\]
where $a_i$ are smooth complex-valued functions of one real variable $x$.

Then, the condition $\{F,H\}=0$ reads
\begin{equation}\label{braFH}
\begin{split}
\{F,H\} &=
h_x{\cdot}\exp(\mu{\cdot}y){\cdot}(h_x{\cdot}a_0(x)_x
-3{\cdot}a_0(x){\cdot}h_{xx}){\cdot}p_x^4
\\
&+h_x{\cdot}\exp(\mu{\cdot}y){\cdot}
(-2\cdot a_1(x){\cdot}h_{xx}+h_x{\cdot}\mu{\cdot}a_0(x)
+h_x{\cdot}a_1(x)_x){\cdot}p_x^3{\cdot}p_y
\\
&+h_x{\cdot}\exp(\mu{\cdot}y){\cdot}
(h_x{\cdot}\mu{\cdot}a_1(x)-3{\cdot}a_0(x){\cdot}h_{xx}
-a_2(x){\cdot}h_{xx}+h_x{\cdot}a_2(x)_x){\cdot}p_x^2{\cdot}p_y^2
\\
&+h_x{\cdot}\exp(\mu{\cdot}y){\cdot}
(h_x{\cdot}\mu{\cdot}a_2(x)+h_x{\cdot}a_3(x)_x
-2\cdot a_1(x){\cdot}h_{xx}){\cdot}p_x{\cdot}p_y^3
\\
&+\tfrac{1}{2}h_x{\cdot}\exp(\mu{\cdot}y){\cdot}
(-a_2(x){\cdot}h_{xx}+h_x{\cdot}\mu{\cdot}a_3(x)){\cdot}p_y^4,
\end{split}
\end{equation} 
where subscripts $a_0(x)_x,h_x$ mean derivation in $x$, and $h_{xx}$ is the
second derivative. Since the monomials $p_x^{4-i}p_y^i$ form a basis of
homogeneous polynomials of degree $4$, every line in \eqref{braFH} should
vanish. This gives us a system of 5 ODEs on 5 functions
$h(x),a_0(x),\ldots,a_3(x)$: each line of \eqref{braFH} corresponds to one ODE.
Subsequently solving the first three of them and resolving $a_3(x)$ from the
last one we obtain
\begin{equation}\label{a0-a3}
\textstyle 
\begin{split}
a_0(x) &\textstyle 
= A_0{\cdot}h_x^3\\
a_1(x) &\textstyle 
= (-\mu{\cdot}A_0{\cdot}h(x)+\frac{A_1}{2{\cdot}\mu}){\cdot}h_x^2\\
a_2(x) &\textstyle 
=  \half{\cdot}(-A_1{\cdot}h(x)
+ \mu^2{\cdot}A_0{\cdot}h(x)^2
+ 3{\cdot}h_x^2{\cdot}A_0+ A_2){\cdot}h_x  \\
a_3(x) &\textstyle 
= \frac{1}{2\mu}{\cdot}(3{\cdot}h_x^2{\cdot}A_0
-A_1{\cdot}h(x)+\mu^2{\cdot}A_0{\cdot}h(x)^2+A_2){\cdot}h_{xx},
\end{split}
\end{equation}
with some constants $A_0,A_1,A_2$. Substituting in the remaining equation
$(\ldots)p_x{\cdot}p_y^3$, we obtain the following non-linear ODE of order $3$ on
$h(x)$:
\begin{equation}\label{Eqh3}
\begin{split}
&(3{\cdot}A_0{\cdot}h_x^2  +\mu^2{\cdot}A_0{\cdot}h(x)^2
-A_1{\cdot}h(x)  +A_2)   {\cdot}h_{xxx}
\;+
\\ + \;& 
6{\cdot}h_{xx}^2{\cdot}h_x{\cdot}A_0+
(6{\cdot}\mu^2{\cdot}A_0{\cdot}h(x) -3{\cdot}A_1 )
      {\cdot}h_x{\cdot}h_{xx}\; +
\\ +\; & 
3{\cdot}\mu^2{\cdot}A_0{\cdot}h_x^3
+(\mu^4{\cdot}A_0{\cdot}h(x)^2
-\mu^2{\cdot}A_1{\cdot}h(x) +\mu^2{\cdot}A_2){\cdot}h_x=0.
\end{split}
\end{equation}
By direct calculations we see  that the equation \eqref{Eqh3} can be written in the form
\begin{equation}\label{Eqh-diff}
\textstyle 
\bigl(\frac{d^2}{dx^2} + \mu^2\bigl)
\bigr( h_x{\cdot}(h_x^2{\cdot}A_0+\mu^2{\cdot}A_0{\cdot}h(x)^2-A_1{\cdot}h(x)+A_2) \bigr)=0.
\end{equation}
Therefore this equation is equivalent to the equation
\begin{equation}\label{Eqh1r}
\textstyle 
 h_x{\cdot}\bigr( h_x^2{\cdot}A_0 +\mu^2{\cdot}A_0{\cdot}h(x)^2 -A_1{\cdot}h(x)+A_2 \bigr)
- \bigl( A_3{\cdot}\frac{\sin(\mu x)}{\mu} +A_4{\cdot}\cos(\mu x) \bigr) =0
\end{equation}
in the sense that $h(x)$ satisfies the equation \eqref{Eqh3} if and only if it
satisfies \eqref{Eqh1r} with the same constant $A_0,A_1,A_2\in \mathbb{C}$ and some constants 
$A_3,A_4\in \mathbb{C}$. Later, in §2.5 (see Theorem \ref{real-h}) we shall show that only real $A_0,A_1,A_2,
A_3,A_4$  are interesting for our purposes.

\subsection{ {\bf Case 2}: $\mu=0$.}  \label{Case2} We proceed as we explained
in §\ref{1.2}: we write the metric in the form
$g=\frac{1}{h_x^2}{\cdot}(dx^2+dy^2)$, so that now
$H=\frac{h_x^2}{2}{\cdot}(p_x^2+p_y^2)$, and then substitute \eqref{a01-mu0} in
the condition $\{F,H\}=0$.  We obtain
\begin{equation}\label{braFHmu0}
\begin{split}
\{F,H\} &=
h_x{\cdot}(h_x{\cdot}a_0(x)_x
-3{\cdot}a_0(x){\cdot}h_{xx}){\cdot}p_x^4
\\
&+h_x{\cdot}
(h_x{\cdot}a_1(x)_x-2{\cdot}a_1(x){\cdot}h_{xx}){\cdot}p_x^3{\cdot}p_y
\\ 
& \textstyle  
+h_x{\cdot} (\half{\cdot}A_1{\cdot}h_x^3 -3{\cdot}a_0(x){\cdot}h_{xx}
-a_2(x){\cdot}h_{xx}+h_x{\cdot}a_2(x)_x){\cdot}p_x^2{\cdot}p_y^2
\\
& \textstyle  
+h_x{\cdot}
(h_x{\cdot}a_3(x)_x
-2{\cdot}a_1(x){\cdot}h_{xx}){\cdot}p_x{\cdot}p_y^3
\\
&+h_x{\cdot}
(A_1{\cdot}h_x^3  + A_3{\cdot}h_x  -2{\cdot}a_2(x){\cdot}h_{xx})){\cdot}p_y^4,
\end{split}
\end{equation} 
with the same constants $A_1,A_3$ as in \eqref{a01-mu0}. This time we can
subsequently resolve all functions $a_0(x),\ldots,a_3(x)$ from the equations and
obtain
\begin{equation}\label{a0-a3mu0old}
\textstyle 
\begin{split}
a_0(x) &\textstyle 
= A_0{\cdot}h_x^3\\
a_1(x) &\textstyle 
= \half{\cdot}\wt A_1{\cdot}h_x^2\\
a_2(x) &\textstyle 
=  \half{\cdot}(3{\cdot}h_x^2{\cdot}A_0  
-A_1{\cdot}h(x) + A_2){\cdot}h_x  \\
a_3(x) &\textstyle 
= \half{\cdot}h_x^2{\cdot}\wt A_1 +\wt A_3,
\end{split}
\end{equation}
with some constants $A_0,\wt A_1,A_2,\wt A_3$. The notation in the formula
above, { especially $\wt A_1,\wt A_3$},   is
chosen for convenience in future formulas. Then the bracket yields
\begin{equation*}\label{braFHmu0subs}
 \{F,H\} =  -h_x^2{\cdot}(3{\cdot}A_0{\cdot}h_x^2{\cdot}h_{xx}
-A_1{\cdot}h_x^2 -A_1{\cdot} h_{xx}{\cdot}h(x)+ A_2{\cdot}h_{xx}-A_3){\cdot}p_y^4.
\end{equation*} 
This means that the equation on $h(x)$ is 
\begin{equation}\label{Eqh2mu0}
3{\cdot}A_0{\cdot}h_x^2{\cdot}h_{xx}-A_1{\cdot}h_x^2 -A_1{\cdot}h_{xx}{\cdot}h(x)+h_{xx}{\cdot}A_2=A_3.
\end{equation} 
The left hand side of this expression is the $x$-derivative of the expression
\begin{equation}\label{lhsEqh1mu0}
h_x{\cdot}(A_0{\cdot}h_x^2 -A_1{\cdot}h(x) + A_2).
\end{equation} 
Therefore the equation \eqref{Eqh2mu0} is equivalent to 
\begin{equation}\label{Eqh1mu0}
h_x{\cdot}(A_0{\cdot}h_x^2 -A_1{\cdot}h(x) + A_2)-(A_3{\cdot}x+A_4)=0.
\end{equation}

\begin{rem} \rm Obviously, we obtain this equation from both the equations
 \eqref{Eqh1all} (i) and (ii) taking the limit $\mu\longrightarrow0$.  Moreover, the solution
 of the Cauchy initial value problem for the equations \eqref{Eqh1r},
 \eqref{Eqh1mu0} depends analytically on all parameters: the variable $x$,
 parameters $\mu,A_0,\ldots,A_4$, the initial point $x_0$, and the initial value
 $h(x_0)$. Therefore we can consider real solutions $h(x)$ of the equation
 \eqref{Eqh1all} as ``real forms'' of a single holomorphic multi-valued
 function $h(x;\mu; A_0,\ldots,A_4; x_0,h_0)$ depending holomorphically on the
 involved parameters. Notice also that the equation \eqref{Eqh1all} (ii) is
 obtained from \eqref{Eqh1all} (i) by replacing $\mu$ by $\cplxi{\cdot}\mu$, and
 similarly for the corresponding cubic integrals. 
\end{rem} 

\begin{rem} \rm The above argumentation shows the existence of one  non-trivial
 cubic integral $F$ in the case $\mu=0$ (i.e., for the metric \eqref{gh} with
 $h$ satisfying (\ref{Eqh1all}(iii)). Namely, such $F$ can be obtained
 substituting the formulas \eqref{a0-a3mu0old} in \eqref{a01-mu0}. The solution $F$ obtained in this way has the form 
 $F=\ti A_3{\cdot}L^3+\ti A_1{\cdot}L{\cdot}H+F_1$ with a fixed cubic integral $F_1$ which
 is linear in $y$.  On the other hand,
 for $\mu\neq0$ in the both cases (i) and (ii) we obtain two non-trivial cubic
 integrals linearly independent of $L^3$ and $L{\cdot}H$, namely, by replacing $\mu$
 by $-\mu$ in formulas \eqref{a0-a3}. It appears that also in the case $\mu=0$ there
 exists another cubic integral $F_2$ that  is (inhomogeneous) quadratic in
 $y$. The latter property is equivalent to the condition $\call^3(F_2)=0$. We
 show the existence of such $F_2$ in the proof of Theorem \ref{thm-4}. {
  This additional integral} $F_2$ is already included in the formulas in
 Theorem \ref{main.th}.

 \smallskip%
 This fact is the reason for the difference in formulas \eqref{a0-a3mu0} and
 \eqref{a0-a3mu0old}. Namely, the substitution of \eqref{a0-a3mu0} in
 \eqref{Fmu0} yields the linear combination $C_1F_1+F_2C_2$ of two cubic
 integrals $F_1,F_2$ which are linear independent of $L^3$ and $L{\cdot}H$.  On
 the other hand, the substitution of \eqref{a0-a3mu0old} in \eqref{a01-mu0}
 yields the linear combination $F=\ti A_3{\cdot}L^3+\ti A_1{\cdot}L{\cdot}H+F_1$ with the
 same $F_1$, which gives only one cubic integral linear independent of $L^3$
 and $L{\cdot}H$

\end{rem}

\subsection{Remaining steps of the proof.} \label{rest-of-proof} %
As we have shown, if a surface metric $g$ admits a linear and a non-trivial
cubic integral $F$, then in appropriate coordinates it has the form
$h_x^{-2}(dx^2+dy^2)$ for some function $h(x)$ satisfying one of the forms
\eqref{Eqh1all} of the Principle equation, and that the cubic integral $F$ can
be constructed using the formula \eqref{a0-a3} or resp.\ \eqref{a0-a3mu0old}.
The remaining steps of the proof are the following:
\begin{itemize}
\item We analyse in which cases the constructed metric $g=h_x^{-2}(dx^2+dy^2)$
 belongs to already known types: Metrics of constant curvature and
 Darboux-superintegrable metrics. This is done in Section \ref{sec3}. We show
 that our metrics are indeed   new examples for most values of the parameters (the values of the parameters corresponding to previously known cases are solutions of certain algebraic equations).
\item In Section \ref{sec4} we prove that in the case of non-constant
 curvature the function $h(x)$ satisfies a \emph{unique up to constant factor}
 equation of type \eqref{Eqh1all}. This result is used to prove the fact that
 if the solution $h(x)$ of the Principle equation with complex parameters
 $\mu;A_0,\ldots,A_4$ is real-valued, then the parameter $\mu$ (which could be apriori
 arbitrary complex number) must be real, purely imaginary, or zero, whereas
 the parameters $A_0,\ldots,A_4$ becomes real  after the application by the appropriate constant. This explains why we
 have only 3 types (i)--(iii) of the Principle equation \eqref{Eqh1all}.
\item In Section \ref{sec:5} we prove that in the case of non-constant
 curvature every type (i)--(iii) of Theorem \ref{main.th} the space of cubic
 integrals has dimension $4$. This means that under hypotheses of the main
 theorem there are exactly 2 non-trivial independent cubic integrals, in addition to $L^3$ and $LH$.  This
 fact in a special case of Kruglikov's ``big gap'' conjecture (see \cite{Kr})
 about possible dimensions of the spaces of cubic integrals of surface metrics.
\end{itemize}

\section{ Special solutions.} 
\label{sec3}

In this section we consider two special  cases of the Principle
equation corresponding to Darboux-superintegrable metrics and constant
curvature metrics.

\subsection{ The case $A_0=0$ corresponds to Darboux-superintegrable
 metrics} \label{A_0=0} %
Recall that a two-dimensional metric $g$ is \slsf{Darboux-superintegrable}, if
the space of its quadratic integrals is at least 4-dimensional and the
curvature is non-constant.  We shall use the following statement which follows
from \cite{Kr} (or even from \cite{Koe}): {\it if a metric $g$ (with the
 Hamiltonian $H$) of non-constant curvature admits a linear integral $L$ and a
 quadratic integral $Q$ such that $L,Q$ and $H$ are functionally independent,
 then $g$ is Darboux-superintegrable.}

This statement implies that { for every real solution $h(x)$ of one of the
 equations\eqref{Eqh1all} with $A_0=0$ the metric $g=h_x^{-2}(dx^2+dy^2)$ is
 Darboux-superintegrable.}

Indeed, { $A_0=0$ if and only if the integral $F$ from Theorem
 \ref{main.th} has zero coefficient at $p_x^3$.} Since the linear integral $L$
 in Theorem \ref{main.th} is $p_y$, the function $Q:=F/p_y$ is an integral
 quadratic in momenta. If $ L, H$ and $F$ are functionally independent, then
 the functions $L,H,Q$ are also functionally independent and the metric is
 Darboux-superintegrable by the result of \cite{Koe,Kr} recalled above.

{ For further use let us note that every Darboux-superintegrable metric  always has the form
 $g=h_x^{-2}(dx^2+dy^2)$ for some function $h(x)$ satisfying one of the
 Principle equations \eqref{Eqh1all} with $A_0=0$. Indeed, for given metric
 $g$ admitting a non-vanishing linear integral $L$ there exists a
 \emph{isothermic} coordinate system $(x,y)$, { unique up to translations},
 in which $L= p_y$. In these coordinates $g$ has the form
 $g=h_x^{-2}(dx^2+dy^2)$ with \emph{some} function $h(x)$. Further, if $Q$ is
 a quadratic integral, then $F:=Q{\cdot}L$ is a cubic integral for $g$. In this
 situation we have shown that $h(x)$ must satisfy one of the equations
 \eqref{Eqh1all} with certain parameters $A_0,\ldots,A_4$ such that
 $F=\sum_{i=0}^3a_i(x,y)p_x^ip_y^{3-i}$ with $a_0=A_0h_x^3$. The condition
 $F=Q{\cdot}L$ means the vanishing of $a_0(x,y)$ which is equivalent to $A_0=0$.}

\subsection{ Parameters in Theorem \ref{main.th}  corresponding to metrics of
 constant curvature.}  
\label{2.7}  %

{ The goal of this  subsection is to understand for what } 
 values of the parameters $A_0,...,A_4$
and the initial value $h(x_0)$ the metric from Theorem \ref{main.th} belong to the previously known classes, that is to the Darboux-integrable metrics and to the metrics of constant curvature. 
In   §\ref{A_0=0} we have shown that  Darboux-superintegrable metrics are characterized
by the condition $A_0=0$.  
Thus in order to understand whether the metrics we constructed are new we need to understand which  metrics with $A_0\ne 0$   have constant curvature. The answer is given in Theorem \ref{R=const}. In particular, 
Corollary \ref{generic-sol}     shows that   most  metrics we constructed  are
  new.

Let $g$ be a metric on $M^2$ of the constant Gauss curvature $R$ and $\mbfv$ a
Killing vector field corresponding to the linear integral $L_\mbfv$. Then
according to the sign of $R$ the Lie algebra of Killing vector fields on $M^2$ is
either $\mathfrak{so}(3)$ (case $R{>}0$), or $\mathfrak{sl}(2,\rr)$ (case $R{<}0$), or
the affine algebra $\mathfrak{aff}(\rr^2)$ of isometries of $\rr^2$ isomorphic to
a semi-direct sum $\mathfrak{so}(2)\ltimes\rr^2$ (remaining case $R=0$).  The
classification of elements of these three Lie algebras gives $6$ types of
Killing vector fields: rotations of $S^2$ ($R{>}0$), rotations, hyperbolic
translations, and loxodromies of the hyperbolic plane ($R{<}0$), and rotations
and translations of $\rr^2$ ($R{=}0$).  Fix a coordinate system $(x,y)$ in
which the metric has the form $g=h_x^{-2}(dx^2+dy^2)$, the curvature is
$R=h_{xxx}{\cdot}h_x-h_{xx}^2$, and the Killing vector field has the form
$\mbfv=\frac{\partial}{\partial y}$. Since in each of these cases the metric
has $3$ Killing vector field, there exists a cubic integral independent of
$L_\mbfv^3$ and $L_\mbfv{\cdot}H$. Consequently, $h(x)$ must satisfy one of the
Principal equations. The explicit situation is as follows:

\begin{thm}\label{R=const} Assume that a  metric  $g=h_x^{-2}(dx^2+dy^2)$ has
 constant Gauss curvature $R$. { Then $h(x)$ satisfies one of the equations
 \eqref{Eqh1all}. Moreover, in this case under additional assumption $A_0\neq0$ 
one of the following possibilities holds:}
\begin{enumerate}
\item[(1)] $h(x)= a{\cdot}\sinh(\mu{\cdot}(x-b)) +c$ with some constants
 $\mu>0,a>0,c,b$ satisfying $R=a^2\mu^4$. In this case the Killing vector field
 $\frac{\partial}{\partial y}$ is locally a rotation of the $2$-sphere of radius
 $r=a\inv\mu^{-2}$ and Gauss curvature $R=a^2\mu^4$. The function  $h(x)$ satisfies the
 equations \eqref{Eqh1all}~\slii (elliptic type) in the form 
\begin{equation}\label{EqRco1}
\begin{split} 
&h_x{\cdot}(h_x^2-\mu^2(h(x)-c)^2+C)= a\mu{\cdot}(C+(a\mu)^2){\cdot}\cosh(\mu{\cdot}(x-b)) \\[2pt]
&h_x{\cdot}(h_x^2-(3\mu)^2(h(x)-c)^2 -3{\cdot}(\mu a)^2 )= -2{\cdot}(a\mu)^3{\cdot}\cosh(3\mu{\cdot}(x-b))
\end{split} 
\end{equation}
with arbitrary constant $C$ in the first equation;
\item[(2)] $h(x)= a{\cdot}\cosh(\mu{\cdot}(x-b)) +c$ with some constants $\mu>0,a>0,c,b$
 satisfying $R=-a^2\mu^4$. In this case the Killing vector field $\frac{\partial}{\partial y}$
 is locally a rotation of the hyperbolic plane of constant Gauss curvature
 $R=-a^2\mu^4$ and $h(x)$ satisfies the equation \eqref{Eqh1all} \slii (elliptic
 type) in the form 
\begin{equation}\label{EqRco2}
\begin{split} 
&h_x{\cdot}(h_x^2-\mu^2(h(x)-c)^2+C)= a\mu{\cdot}(C-(a\mu)^2){\cdot}\sinh(\mu{\cdot}(x-b)) \\[2pt]
&h_x{\cdot}(h_x^2-(3\mu)^2(h(x)-c)^2 +3{\cdot}(\mu a)^2 )= -2{\cdot}(a\mu)^3{\cdot}\sinh(3\mu{\cdot}(x-b))
\end{split} 
\end{equation}
with arbitrary constant $C$ in the first equation;
\item[(3)] $h(x)= a{\cdot}\sin(\mu{\cdot}(x-b)) +c$ with some constants $\mu>0,a>0,c,b$
 satisfying $R=-a^2\mu^4$. In this case the Killing vector field $\frac{\partial}{\partial y}$
 is locally a translation on the hyperbolic plane of constant Gauss curvature
 $R=-a^2\mu^4$ and $h(x)$ satisfies the equation \eqref{Eqh1all} \sli
 (hyperbolic type) in the form 
\begin{equation}\label{EqRco3}
\begin{split} 
&h_x{\cdot}(h_x^2+\mu^2(h(x)-c)^2+C)= a\mu{\cdot}(C+(a\mu)^2){\cdot}\cos(\mu{\cdot}(x-b)) \\[2pt]
&h_x{\cdot}(h_x^2+(3\mu)^2(h(x)-c)^2 -3{\cdot}(\mu a)^2 )= 2{\cdot}(a\mu)^3{\cdot}\cos(3\mu{\cdot}(x-b))
\end{split} 
\end{equation}
with arbitrary constant $C$ in the first equation;
\item[(4)] $h(x)= a{\cdot}(x-b)^2+c$ with some constants $a>0,c,b$
 satisfying $R=-4a^2$. In this case the Killing vector field $\frac{\partial}{\partial y}$
 is  a loxodromy on the hyperbolic plane of constant Gauss curvature
 $R=-4a^2$ and $h(x)$ satisfies the equation \eqref{Eqh1all} \sliii
 (parabolic/nilpotent type) 
\begin{equation}\label{EqRco4}
h_x{\cdot}(h_x^2-4{\cdot}a{\cdot}h(x)+A_2)=2a{\cdot}(A_2-4{\cdot}a{\cdot}c){\cdot}(x-b)
\end{equation}
with arbitrary constant $A_2$;
\item[(5)] $h(x)= a{\cdot}\exp(\mu x) +c$ with some constants
 $\mu>0,a>0,c$ and $R=0$. In this case the Killing vector field
 $\frac{\partial}{\partial y}$ is locally a rotation of the Euclidean plane ($R=0$) and
 $h(x)$ satisfies the equation \eqref{Eqh1all} \sli (hyperbolic type)
in the form 
\begin{equation}\label{EqRco5}
\begin{split} 
&h_x{\cdot}(h_x^2-\mu^2(h(x)-c)^2+C)=- a\mu C{\cdot}\exp(\mu x) \\[2pt]
&h_x{\cdot}(h_x^2-(3\mu)^2(h(x)-c)^2 )= -8{\cdot}(a\mu)^3{\cdot}\exp(3\mu x)
\end{split} 
\end{equation}
with arbitrary constant $C$ in the first equation;
\item[(6)] $h(x)= a{\cdot}x+c$ with some constants $a>0,c$.   In
 this case  $R=0$, the Killing vector field $\frac{\partial}{\partial y}$ is locally a translation
 of the Euclidean plane ($R=0$), and $h(x)$ satisfies the equation
 \eqref{Eqh1all} \sliii (parabolic/nilpotent type) 
\begin{equation}\label{EqRco6}
h_x{\cdot}(h_x^2-A_1{\cdot}h(x)+A_2)=-a^2A_1{\cdot}x+a{\cdot}(a^2-c{\cdot}A_1+A_2)
\end{equation}
with arbitrary constants $A_1,A_2$.
\end{enumerate} 
\end{thm}

\proof As we have shown above, if a metric $g$ admits a Killing vector field
$\mbfv$, then in appropriate coordinates $g$ has the form
$g=h_x^{-2}(dx^2+dy^2)$ and the Killing vector field the form
$\mbfv=\frac{\partial}{\partial y}$. In this case the Gauss curvature $R$ is given by
$R=h_{xxx}{\cdot}h_x-h_{xx}^2$. Thus we are interesting in possible solutions of
the ODE $h_{xxx}{\cdot}h_x-h_{xx}^2=R$ with constant parameter $R$ such that
$h_x\neq0$. By direct calculations we see that  every function on the list items  $(1)$--$(6)$ satisfies the ODE
$h_{xxx}{\cdot}h_x-h_{xx}^2=R$ with an appropriate constant $R$, and the theorem
claims that the list is complete. In view of the uniqueness of the solution of
an ODE with the given initial values we must show that every combination of
the initial values $I:=(R,x_0,h(x_0),h_x(x_0),h_{xx}(x_0))$ is realised by
one of the solution on the list.  Inverting the sign of $h(x)$ and $x$, if
needed, we may assume that $h_x(x_0)>0$ and $h_{xx}(x_0)>0$.

\smallskip %
Let us consider $h_{xxx}(x_0)=\frac{R+h_{xx}(x_0)^2}{h_{x}(x_0)}$. If
$h_{xxx}(x_0)=0$ then the data $I$ are realised by an appropriate polynomial
of degree $2$ (case $h_{xx}(x_0)\neq0$, list item (4)\,) or $1$ (case
$h_{xx}(x_0)=0$, list item (6)\,).

In the case $h_{xxx}(x_0)<0$ we set $\mu:=\sqrt{- h_{xxx}(x_0)/h_x(x_0)}$ and
$a:= \sqrt{ \mu^{-2}h_x^2(x_0)+ \mu^{-4}h_{xx}^2(x_0)}$. It is not difficult to
see that the ODE $h_{xxx}{\cdot}h_x-h_{xx}^2=R$ admits the solution
$h(x)=a{\cdot}\sin(\mu{\cdot}(x-b)) +c$ (list item (3)\,) with appropriate parameters
$b$ and $c$ satisfying the initial conditions $I$.

In the remaining case $h_{xxx}(x_0)>0$ we set
$\mu:=\sqrt{h_{xxx}(x_0)/h_x(x_0)}$ and look for the solution of the equation
$h_{xxx}{\cdot}h_x-h_{xx}^2=R$ in one of the forms (1), (2), or (5) with
appropriate parameters $a>0,b,c$. The form (1)
is realised in the case $h_{xx}(x_0)<\mu{\cdot}h_{x}(x_0)$ in which $R>0$, the form
(2) in the case $h_{xx}(x_0)>\mu{\cdot}h_{x}(x_0)$ in which $R>0$, and form (5) in
the case $h_{xx}(x_0)=\mu{\cdot}h_{x}(x_0)$ in which $R=0$. The needed parameters
$a>0,b,c$ can be found easily.

\medskip %
It remains to show every function $h(x)$ given by one of the formulas (1)--(6)
satisfies one of the ODEs \eqref{Eqh1all} with $A_0\neq0$ and determine possible
values of the parameters $\mu$ and $A_0,\ldots,A_4$. Due to the condition $A_0\neq0$ we
may assume that $A_0=1$. The key observation is that $h(x)$ is (up to a
constant) either a trigonometric (case (3)\,), or trig-hyperbolic (cases (1)
and (2)\,), or exponential (case (5)\,), or a usual monomial (cases (4) and
(6)\,) and therefore the differential expression
$h_x{\cdot}(h_x^2±\ti\mu^2{\cdot}h^2(x)-A_1{\cdot}h(x)+A_2)$ will be a polynomial of the
same type, divisible by the monomial $h_x$, for example, $\sum_jB_je^{j\mu x}$ in
the exponential case.

In the cases (1) and (2) we conclude that the right hand side  must be of the form
$B_1{\cdot}\cosh(k\mu)+B_2{\cdot}\sinh(k\mu)$ with $k=1,2$ or $3$ which gives
$\ti\mu=k\mu$. The case $k=2$ is excluded by the argument that for $k\neq±1$ the
expression $h_x{\cdot}(h_x^2±\ti\mu^2{\cdot}h^2(x))=h_x{\cdot}(h_x^2± k^2\mu^2{\cdot}h^2(x))$ is a
trig-hyperbolic polynomial of degree $3$, i.e., containing a term $\cosh(3\mu)$
or a term $\sinh(3\mu)$. Using the relations
$\sinh(3x)=4{\cdot}\sinh^3(x)+3{\cdot}\sinh(x)$ and
$\cosh(3x)=4{\cdot}\cosh^3(x)-3{\cdot}\cosh(x)$ we conclude that the only possible
equations are \eqref{EqRco1} and \eqref{EqRco2}.

The remaining case (3)--(6) involving trigonometric polynomials,
exponential polynomials, and usual polynomials instead of
trig-hyperbolic ones are treated  in the same manner. \qed

\begin{corol}\label{generic-sol}
 Every equation \eqref{Eqh1all} with $A_0\neq0$ and $(A_3, A_4)\neq(0,0)$ admits only
 finitely many (real) solutions $h(x)$ such that the metric
 $g=h_x^{-2}(dx^2+dy^2)$ has constant Gauss curvature, except the case of the
 equation $h_x(A_0{\cdot}h_x^2+A_2)=A_4$ which always admits a solution of the
 form $h(x)= a{\cdot}x+c$ with arbitrary $c$ and $a$ satisfying
 $a(A_0{\cdot}a^2+A_2)=A_4$.
\end{corol}

Every real solution $h(x)$ of \eqref{Eqh1all} is completely determined by its
initial values $h(x_0),h_x(x_0)$ at a given point $x_0$. Thus for a generic
choice of the initial value $h(x_0)$ the solution $h(x)$ of the equation
\eqref{Eqh1all} with this initial value and with any root $h_x(x_0)$ of the
corresponding algebraic equation at $x_0$ the metric $g=h_x^{-2}(dx^2+dy^2)$
has non-constant Gauss curvature.

\proof  As we have seen,
a metric of the form $g=h_x^{-2}(dx^2+dy^2)$ has constant curvature if and
only if $h(x)$ is one of the forms (1)--(6). Let us consider possible right
hand sides.

Every expression $A_3{\cdot}\frac{\sin(\mu{\cdot}x)}{\mu} +A_4{\cdot}\cos(\mu{\cdot}x)$ can be
written in the form $A{\cdot}\cos(\mu{\cdot}(x-b))$ with unique $A$ and $b$ unique up to
a multiple of the period. Similarly, every expression
$A_3{\cdot}\frac{\sinh(\mu{\cdot}x)}{\mu}\allowbreak+A_4{\cdot}\cosh(\mu{\cdot}x)$ can be uniquely
written in one of the following forms:\break $A{\cdot}\cosh(\mu{\cdot}(x-b))$,
$A{\cdot}\sinh(\mu{\cdot}(x-b))$, $A{\cdot}\exp(\mu{\cdot}x)$, or $A{\cdot}\exp(-\mu{\cdot}x)$. The latter
case can be reduced to the previous one by inverting the $x$-axis.  Thus the
right hand side of the equation \eqref{Eqh1all}, \sliip, determines which type
(1), (2), or (5) of the solution $h(x)$ we obtain, and in the case
\eqref{Eqh1all}, \slip, the solution must be of the type (3).

In the case when $h(x)$ is a solution of the type (1), (2), or (3) we proceed
as follows: Comparing the right hand side of equations \eqref{Eqh1all} and
\eqref{EqRco1}--\eqref{EqRco3} we determine $b$ and possible values of $\mu$,
there are only finitely many such possibilities. Then multiplying the equation
by a constant we make $A_0=1$. Next, we compare the l.h.s.\ and determines the
parameters $c$ and $C$. After this the right hand side of
\eqref{EqRco1}--\eqref{EqRco3} determines the possible values of $a$.  Clearly,
we have only finitely many possibilities.

Notice that the type (5) is not generic itself since it occurs only if
$A_3=±\mu{\cdot}A_4$. Nevertheless, in this case for a given $A_0,\ldots,A_4$ we still
have only finitely many solutions $h(x)$ giving constant curvature. Indeed, we
determine possible values of $\mu$ considering the right hand side of the
equation, then from l.h.s.\ we determines possible values of the parameters
$c$ and $C$, and finally again from the right hand side we determine $a$.

In the case (4) when $h(x)= a{\cdot}(x-b)^2+c$ we must have $b=-A_4/A_3$ and
$a=A_1/4$, and finally $c= \frac{2aA_2- A_3}{8a^2}$. So for given $A_0,\ldots,A_4$
we could have at most one solution of type (4).

Finally, if $A_1\neq0$ and $h(x)$ is of type (6), i.e.\ $h(x)= a{\cdot}x+c$, then $a$
must satisfy $A_3=-a^2{\cdot}A_1$ which gives us at most two possibilities. 
For every $a$ we have the unique possibility for $c$.
\qed

\section{ Uniqueness
of the Principle equation } 
\label{sec4}

The uniqueness of the Principle equation is an interesting phenomenon {\it per
 se} and plays an important role in the proof of the main theorem.
We shall need the following two results. 

\begin{lem}\label{asymp}
 Let $h(x)$ be a complex-valued solution of the equation 
\begin{equation}\label{Eq-asymp}
\cale:= h_x(h_x^2-9{\cdot}h(x)^2+A_2)- A_+e^{3x}-A_-e^{-3x}=0
\end{equation}
with complex coefficients $A_+\neq0,A_-,A_2$ defined for $x\in[x^*,+\infty)$. Assume
that for $x\to+\infty$ the function $h(x)$ has the asymptotic growth
$h(x)=a{\cdot}e^x+o(e^x)$ with $a\neq0$. Then $A_+=-8a^3$ and there exists a complex-valued
real-analytic function $f(\tau)$ defined for sufficiently small $\tau$ such that
$f(0)=a$ and $h(x)=e^x{\cdot}f(e^{-2x})$.
\end{lem}

\proof Write $h_x(x)=\psi(x){\cdot}e^{x}$, substitute this expression in
\eqref{Eq-asymp}, and consider the obtained relation as a cubic algebraic
equation on a variable $\psi$ depending on the parameter $x$. Then for $x\to+\infty$
(the coefficients of) the obtained equation converge to
$\psi(\psi^2-9a^2)=A_+$. This implies the asymptotic growth $h_x(x)=a'{\cdot}e^x+o(e^x)$
with some $a'$ satisfying the equation $a'(a'{}^2-9a^2)=A_+$. Integrating it we
obtain the asymptotic $h(x)=a'{\cdot}e^x+o(e^x)$. Consequently, $a'=a$ and hence
$a$ satisfies $A_+=-8a^3$. 

Now make the substitution $x=-\half\log(\tau)$ and
$h(x)=e^x{\cdot}(a+ a_0e^{-2x}+ e^{-4x}f(e^{-2x}))=\tau^{-1/2}(a+a_0\tau+\tau^2f(\tau))$. Then 
the equation \eqref{Eq-asymp} transforms into
\begin{equation}\label{Eqf-tau1}
\begin{split} 
8f_\tau^3\,\tau^8  +36ff_\tau^2\,\tau^7+
12f_\tau(3f^2+f_\tau a_0)\,\tau^6-12af_\tau^2\,\tau^5 & \\
-12(3a_0f^2 + 6ff_\tau a+a_0^2f_\tau)\tau^4 
+(2A_2f_\tau-48aa_0f_\tau-72af^2-36a_0^2f)\tau^3& \\
+(3A_2f+A_--72aa_0f-8a_0^3-12a^2f_\tau)\tau^2& \\
+(A_2-12aa_0)(a_0\tau -a)
& =0. 
\end{split}
\end{equation}
This means that we are now looking for solutions $f(\tau)$ of \eqref{Eqf-tau1}
defined for small $\tau>0$. The condition on the growth of $h(x)$ and $h_x$ means
that $f(\tau)=o(\tau^{-3/2})$ and $f_\tau(\tau)=o(\tau^{-2})$. Therefore we must have
$a_0=\frac{A_2}{12a}$ and the substitution $A_2=12aa_0$ transforms the equation
\eqref{Eqf-tau1} into
\begin{equation}\label{Eqf-a0}
\begin{split} 
8f_\tau^3\,\tau^6  +36ff_\tau^2\,\tau^5+
12f_\tau(3f^2+f_\tau a_0)\,\tau^4-12af_\tau^2\,\tau^3 & \\
-12(3a_0f^2 +6ff_\tau a + a_0^2f_\tau)\tau^2 
-12(2aa_0f_\tau +6af^2+3a_0^2f)\tau& \\
+(A_-  -12a^2f_\tau  -36aa_0f -8a_0^3)
& =0. 
\end{split}
\end{equation}

For any given $A_-,a,a_0,f$ and sufficiently small $\tau$ the latter relation can
be resolved in $f_\tau$ as a real-analytic function $f_\tau=F(\tau,f,a,a_0,A_-)$ with 
$F(\tau,f,a,a_0,A_-)=\frac{A_-  -36aa_0f(\tau) -8a_0^3}{12a^2}+ O(\tau)$.

Consequently, we can conclude the following properties: Any solution $h(x)$
of \eqref{Eq-asymp} satisfying the hypotheses of the lemma is given by the
series $h(x)=ae^x +\sum_{k=0}^\infty a_ke^{-(2k+1)x}$ which converges for
$x\in[x_0,+\infty)$. Moreover, the coefficients $a,a_0$ satisfy the relations
$8a^3=-A_+$ and $12aa_0=A_2$. 

Furthermore, we can also conclude the following two existence results for
solutions of \eqref{Eq-asymp}:

$\bullet$ \ First, for $a,a_0$ satisfying the conditions above, for any given
sufficiently large $x_0\gg0$ and any sufficiently small $b\in\cc$ there exists a
unique solution of \eqref{Eq-asymp} with the initial value
$h(x_0)=a^{x_0}+a_0e^{-x_0}+be^{-3x_0}$.

$\bullet$ \ Second, for $a,a_0$ satisfying the conditions above and any given $a_1$
there exists a unique solution of \eqref{Eq-asymp} which is defined for $x\gg0$
and whose initial terms in the series above have coefficients $a,a_0,a_1$.
\qed

\begin{lem}\label{rhs=0} Assume that a complex-valued function $h(x)$
 satisfies the equation
\[
h_x(A_0(h_x^2-\mu^2h(x)^2) -A_1\,h(x) +A_2)=0.
\]
with complex parameters $\mu,A_0,A_1,A_2$.  Then $R:=h_{xxx}h_x-h_{xx}^2$ is
constant.
\end{lem}

\proof We have obviously two possibilities: Either $h_x$ vanishes identically,
$h_x\equiv0$, or
\begin{equation}\label{Eq-Lmu}
A_0(h_x^2-\mu^2h(x)^2) -A_1\,h(x) +A_2\equiv0.
\end{equation}
The first case $h_x\equiv0$ is trivial since then $R=h_{xxx}h_x-h_{xx}^2\equiv0$. So we
may assume that $h_x$ is not vanishing identically. 

Assume additionally that $\mu\neq0$. In the
case $A_0=0$ the solution of \eqref{Eq-Lmu} is a constant function, and $R$
vanishes also. In the case $A_1=A_2=0$ the solution
of \eqref{Eq-Lmu} is $h(x)=C{\cdot}\exp(±\mu x)$, and again $R$ vanishes. In the
remaining case $A_0\neq0\neq A_1$ every solution of \eqref{Eq-Lmu} has the form 
$h(x)=c_0+c_1\sinh(±\mu x+c_2)$ with arbitrary $c_2$ and appropriate
$c_0,c_1$. This time $R$ must be constant too.

In the case $\mu=0$ the argumentation is changed as follows. If $A_0=0$,
then $h(x)$ must be constant, and then $R\equiv0$. If $A_1=0$, then
$h(x)$ must be linear, which is also a contradiction. Finally, in the case
$A_0\neq0\neq A_1$ every solution $h(x)$ of \eqref{Eq-Lmu} with $\mu=0$ is
quadratic in $x$, and then $R:=h_{xxx}{\cdot}h_x-h_{xx}^2$ is constant again. 
\qed

\begin{thm}\label{thm-uniq} Let $h(x)$ be a complex function defined in a
 some open set $U\subset\cc$ satisfies two equations each of the form \eqref{Eqh1all}
 with some complex parameters $A_0,\ldots,A_4$, $\mu$ and
 respectively $B_0,\ldots,B_4$, $\lambda$. Assume that
 $R:=h_{xxx}{\cdot}h_x-h_{xx}^2$ is not constant. Then $\mu=±\lambda$ and the equations
 are proportional.
\end{thm}

\begin{rem} \rm 
 For convenience in calculation below we consider only equations of
 the form \eqref{Eqh1all} (ii) or (iii), but not \eqref{Eqh1all} (i). This is
 an equivalent problem, since the substitution $\mu\mapsto\cplxi\mu$ switches between
 forms \eqref{Eqh1all} (i) and (ii).
\end{rem}

\proof One of the techniques to prove the theorem is to write a Taylor series
$h(x)=\sum_ja_j(x-x_0)^j$, substitute it in both equations, write the expansions,
and then compare term by term the coefficients. In some places we use another
 approach, namely, we study geometric properties of the
solution $h(x)$ using methods of geometric function theory and algebraic
geometry. It should be noticed however that every relation which will be
obtained by geometric methods can be also received purely
algebraically from the equations obtained from Taylor series.

\medskip %
Denote by $\cale_\lambda$ and $\cale_\mu$ the equations from the hypotheses of the
theorem, and $\Lambda_\mu(x)$, $\Lambda_\lambda(x)$ the r.h.s.-s of these equations. In the case
$\mu\neq0$ and resp.\ $\lambda\neq0$ we rewrite them as $\Lambda_\mu=A_+e^{\mu x}+A_-e^{-\mu x}$ and
resp.\ $\Lambda_\lambda=B_+e^{\lambda x}+B_-e^{-\lambda x}$. Notice that by Lemma \ref{rhs=0} $\Lambda_\lambda(x)$
and $\Lambda_\mu$ are non-zero.

\smallskip %
We claim that the function $h(x)$ extends to an analytic multi-sheeted ($\Leftrightarrow$
multi-valued) function of $x\in\cc\bs S$ for some discrete set $S\subset\cc$. Let us
consider several cases. The first is when $A_0\neq0$. In this case the equation
$\cale_\mu$ is a polynomial of degree $3$ in $h_x$. Let $\cald_\mu(x)$ be its
discriminant with respect to $h_x$. Then
$\cald_\mu(x)=-27\Lambda_\mu^2-4(A_0\mu^2h(x)^2-A_1h(x)+A_2)^3$, this is an analytic
function in $x$ defined in the domain of definition $U$ of $h(x)$. In the case
when $\cald_\mu(x)$ is not identically zero we can resolve the equation
$\cale_\mu$ as an analytic multi-sheeted ($\Leftrightarrow$ multi-valued) function of %
$x\in U\bs S$ where $S$ is the set of zeroes of $\cald_\mu(x)$. This transforms the
equation in the explicit form $h_x=F(x,h(x))$ where $F$ is multi-sheeted
function with ramifications exactly at zero points of the discriminant
$\cald_\mu$. This gives us the claim. 

In the case when the discriminant $\cald_\mu$ vanishes identically $h_x$
satisfies also the equation $3h_x^2+A_0\mu^2h(x)^2-A_1h(x)+A_2=0$ which is the
derivative of $\cale_\mu$ with respect to $h_x$. Again we obtain an explicit
equation $h_x=F(x,h(x))$ with analytic multi-sheeted right hand side
$F(x,h(x)$ with singularities in some discrete subset $S$.  Hence this time
also $h(x)$ extends to an analytic multi-sheeted function of $x\in\cc\bs S$ for
some discrete set $S\subset\cc$. Finally, in the case $A_0=0$ the equation $\cale_\mu$
can be resolved as $h_x=\frac{\Lambda_\mu(x)}{A_2-A_1h(x)}$, and we can conclude the 
claim.

\smallskip%
Assume that  $A_0\neq0\neq B_0$. Then dividing equations by $A_0$ or
resp.\ $B_0$ we reduce the general situation to the case $A_0=B_0=1$.

\smallskip %
First, we prove that under the hypotheses of the theorem we must have
the relation $\lambda=±\mu$.  Let us assume the contrary, i.e.,
$\lambda\neq±\mu$. We shall consider numerous special cases and subcases.

Without loss of generality we may suppose that $|\lambda|\geq|\mu|$. In
particular, $\lambda\neq0$. Observe that under complex affine transformation $x\mapsto ax+b$
the equations $\cale_\lambda,\cale_\mu$ retain their structure only changing the
parameters $\mu,\lambda,A_i,B_i$, in particular, $\lambda$ transforms in $\frac{\lambda}{a}$.
Consequently, we may assume that $\lambda$ is real positive, $\lambda>0$. Then by
assumption on $\mu$ we conclude $\big|\Re(\mu)\big|<\lambda$. Further, recall that
$\Lambda_\lambda(x)=B_+{\cdot}\exp(\lambda x) + B_-{\cdot}\exp(-\lambda x)$ such that at least one constant
$B_+,B_-$ is non-zero. Inverting the coordinate $x$, if needed, we can
suppose that $B_+\neq0$.

Assume additionally that $\mu\neq0$.  Then $\Lambda_\mu(x)=A_+{\cdot}\exp(\mu x)+A_-{\cdot}\exp(-\mu
x)$. %
Consider the difference $\cale_\lambda-\cale_\mu$. It has the form
\begin{equation}\label{eq-diff}
h_x{\cdot}(C_0 h(x)^2 + C_1 h(x) + C_2) =
\Lambda_\delta(x)
\end{equation}
where $C_0=\lambda^2-\mu^2$ and $\Lambda_\delta(x)=\Lambda_\lambda(x)-\Lambda_\mu(x)$. Denote this equation by
$\cale_\delta$.  Integrating it, we obtain an \emph{algebraic} equation
\begin{equation}\label{eq-int}
\textstyle
\frac13 C_0 h(x)^3 + \frac12 C_1 h(x)^2 + C_2h(x)=\wt \Lambda_\delta(x)+ C_3
\end{equation}
which we denote by $\wt\cale_\delta$ and in which $C_3$ is some
constant and $\wt \Lambda_\delta(x)=\int \Lambda_\delta(x)dx$ equals
\begin{equation}\label{ti-L-del}
\textstyle
\wt \Lambda_\delta(x)= \frac{B_+}{\lambda}{\cdot}\exp(\lambda x) 
- \frac{B_-}{\lambda}{\cdot}\exp(-\lambda x) + 
\frac{A_+}{\mu}{\cdot}\exp(\mu x) 
- \frac{A_-}{\mu}{\cdot}\exp(-\mu x). 
\end{equation}

Making a translation in $x$, we can suppose that the real axis $x\in\rr$ does
not contain singular points of $h(x)$, and that our open set $U\subset\cc$ hits the
real axis $x\in\rr$. Then there exists a unique extension of the function $h(x)$
over the axis $x\in\rr$ which satisfies the equation \eqref{eq-int}.

Since $C_0=\lambda^2-\mu^2\neq0$, the function $h(x)$ has asymptotic expansion
$h(x)\sim\Big(\frac{3B_+}{C_0\lambda}\Big)^{1/3}\allowbreak\exp(\lambda x/3)$ for
$x\longrightarrow+\infty$. From \eqref{eq-diff} we conclude that the derivative $h_x$ has
asymptotic expansion
$h(x)\sim\Big(\frac{3B_+}{C_0\lambda}\Big)^{1/3}{\cdot}\frac{\lambda}{3}{\cdot}\exp(\lambda x/3)$ for
$x\longrightarrow+\infty$.

Further, assume that $\lambda\neq±3\mu$. Then $h_x^2-\mu^2h(x)^2$ has asymptotic expansion
$\sim C{\cdot}\exp(2\lambda x/3)$ for $x\longrightarrow+\infty$, which gives the asymptotic expansion $\sim
C{\cdot}\exp(\lambda x)$ for the r.h.s.\ of $\cale_\mu$. But by our condition
$\big|\Re(\mu)\big|<\lambda$ the l.h.s.\ $\Lambda_\mu(x)$ has slower growth. The obtained
contradiction gives the proof in the case $\lambda\neq±3\mu$.

It remains to consider the case $\lambda=±3\mu$. However, before this case we notice
that the consideration above are valid also in the case $\mu=0\neq\lambda$. Indeed, we
obtain the same growth asymptotic
$h(x)\sim\Big(\frac{3B_+}{C_0\lambda}\Big)^{1/3}{\cdot}\frac{\lambda}{3}{\cdot}\exp(\lambda x/3)$ for
$x\longrightarrow+\infty$, and the same contradiction in growth of right and left hand sides of 
$\cale_\mu$. 

Now we consider the case $\lambda=±3\mu$. Recall that we assume that $A_0\neq0\neq B_0$.
Changing the sign of $\mu$, if needed, we obtain $\lambda=+3\mu$. Rescaling $x$, we can
make $\lambda=3$ and $\mu=1$. Adding to $h(x)$ a constant we may assume that
$B_1=0$. At this point we apply Lemma \ref{asymp} to $h(x)$ and the equation
$\cale_\lambda$. It gives us the presentation of $h(x)$ as a series
$h(x)=ae^x+a_0e^{-x}+a_1e^{-3x}+\cdots$ which converges for $x\gg0$ such that
$B_+=-8a^3$ and $12aa_0=B_2$. Differentiating the series we obtain
$h_x=ae^x-a_0e^{-x}-3a_1e^{-3x}+\cdots$. This gives us
$h_x^2-h(x)^2=(h_x-h(x)){\cdot}(h_x+h(x))=-4aa_0-8aa_1e^{-2x}+\cdots$. In the case
$A_1\neq0$ the growth of the l.h.s.\ of $\cale_\mu$ is $\sim e^{2x}$, which is faster
than in the r.h.s.\ $\Lambda_\mu\sim e^x$. Consequently, $A_1=0$.

Consider the case $B_-\neq0$. Then repeating the argumentation above, we obtain
the asymptotic behaviour $h(x)\sim a_-e^{-x}$ for $x\to-\infty$. After this we can apply
Lemma \ref{asymp} to $h(x)$ and the equation $\cale_\lambda$ in the negative range
$x\in(-\infty,x_0]$, yielding a similar expansion
$h(x)=a_-e^{-x}+a_{0,-}e^{x}+a_{1,-}e^{3x}+\cdots$, in which $B_-=8a_-^3$ and
$12a_-a_{0,-}=B_2$. Under translation   $x\mapsto x+\xi$ with $\xi\in\cc$ the coefficients
$a,a_-$ transform as $a\mapsto ae^{-\xi},a_-\mapsto a_-e^{+\xi}$. Consequently after an
appropriate translation we achieve the equality $a=-a_-$, and hence the
equality $B_+=-8a^3=8a_-^3=B_-$. The relations $B_2=12aa_0=12a_-a_{0,-}$
gives also the equality $a_{0,-}=-a_0$.

The substitution of the series in $\cale_\mu$ yields
$(aA_2-A_+-4a^2a_0)e^x+O(e^{-x})$ for $x\to+\infty$ and
$(-a_-A_2-A_-+4a_-^2a_{0,-})e^{-x} +O(e^{x})$ for $x\to-\infty$. Since $\cale_\mu$
vanishes identically, we obtain $A_+=aA_2-4a^2a_0=-a_-A_2-4a_-^2a_{0,-}=A_-$.
The next term in the expansion of $\cale_\mu$ for $x\to+\infty$ is
\[
(-8a^2a_1+4a_0^2a-a_0A_2+4a^2a_0-aA_2)e^{-x}.
\]
Since it must vanish, we obtain
$a_1=-(aA_2-4a_0^2a+a_0A_2-4a^2a_0)/(8a^2)$. Making the same computation for
$x\to-\infty$ we obtain $a_{1,-}=-a_1$.

We conclude that the solution $h(x)$ has sheets which satisfy the relation
$h(-x)=-h(x)$ Further, recall that $h(x)$ satisfies the algebraic equation
$\ti\cale_\delta$, see \eqref{eq-int}. Since $A_+=A_-$ and $B_+=B_-$, $\ti \Lambda_\delta$
(given by \eqref{ti-L-del} with $\lambda=3$ and $\mu=1$) is an odd function, $\ti
\Lambda_\delta(-x)=-\ti \Lambda_\delta(x)$. Consequently, $C_3$ vanishes.

Let $w(z)$ be the 3-sheeted function of the argument $z\in\cc$ given by the
algebraic equation $\frac83w^3+C_2w=z$ with $C_2\neq0$. Then %
$h(z)=w(\ti \Lambda_\delta(x))$. Let us observe the following facts about the function
$w(z)$: The first is that or $z$ small enough three branches of $w(z)$ are
given by the approximate formulas $w_0(z)=z/C_2+O(z^2)$,
$w_+(z)=+\sqrt{-3C_2/8}$, and $w_-(z)=-\sqrt{-3C_2/8}$. The second is that
the monodromy group of $w(z)$ is the symmetric group permuting the sheets of
$w(z)$. In particular going along an appropriate path in $z$-plane we can
interchange two given sheets of $w(z)$. 

In turn, the function $z=\wt \Lambda_\delta(x)$ is the composition of the polynomial
$Q(u):=\frac83 B_+u^3 +2(B_+-A_+)u$ with the functions $u=\sinh(x)$. The
polynomial $Q(u)$ has the same structure $c_0z^3+c_2z$ as the polynomial
$\frac83w^3+C_2w$. Consequently, there are two possibilities: Either the
critical values of the polynomials $Q(u)$ and $\frac83w^3+C_2w$ are different,
and then monodromy of the composition $w(Q(u))$ in the full symmetric group as
in the case of the function $w(z)$, or the difference $\frac83w^3+C_2w-Q(u)$
splits in the product $\frac83\prod_{j=1}^3(w-b_ju)$ with appropriate
$b_j\in\cc$. In the latter case three possible branches of $h(x)$ are
$w(u(\sinh(x)))=b_j\sinh(x)$, and in this case $R=h_{xxx}h_x-h_{xx}^2$ is
constant in contradiction with the assumption of the theorem. Consequently,
the latter case is impossible, and the monodromy of the composition $w(Q(u))$
is the full symmetric group.

Finally, the critical points of the function $u=\sinh(x)$ are given by the
condition $\sinh'(x)=\cosh(x)=0$, and hence the critical values of
$u=\sinh(x)$ are $±\cplxi$. The corresponding values of $z=Q(u(x))$ are
$z_±:=±\cplxi(\frac23B_++2A_+)$. Here we notice that for each value $z_±$ has
at least two pre-images $Q\inv(z_±)$ such that at most one of them is
$±\cplxi$, and neither of these pre-images is $0$. Further, we notice that the
function $u=\sinh(x)$ is surjective. Indeed, it is the composition of the
rational function $u(t)=\frac12(t-t\inv)$ and the exponent $t=\exp(x)$, the
map $u(t)=\frac12(t-t\inv)$ acts surjectively from $\cc\bs\{0\}$ onto $\cc$, the
map $t=\exp(x)$ surjectively from $\cc$ onto $\cc\bs\{0\}$. Summing up we
conclude that for any of two critical values $z^*$ of the polynomial
$\frac83w^3+C_2w$ there exists an $x^*\in\cc$ which is not a critical point of
the function $z=Q(\sinh(x))$ but $z^*=Q(\sinh(x^*))$. This shows that the
monodromy of the function $h(x)$ is the full symmetric group permuting its 3
sheets($\Leftrightarrow$ branches). 

\smallskip %
In particular, each of 3 sheets of $h(x)$ satisfy both equations $\cale_\mu$ and
$\cale_\lambda$. Notice that since the l.h.s.\ $\ti \Lambda_\delta(x)$ of \eqref{eq-int}
vanishes at $x=0$, there exists a branch $h(x)$ which vanishes at
$x=0$. Moreover, its behaviour is $h(x)=\ti \Lambda_\delta(x)/C_2+O(\ti \Lambda_\delta^2(x))$.
Further, $\ti \Lambda_\delta(x)=2(B_+-A_+)x+O(x^3)$ in the case $B_+\neq A_+$, and 
$\ti \Lambda_\delta(x)=\frac83B_+x^3+O(x^5)$ otherwise. In any case, this branch $h(x)$
is regular at $x=0$. 

Now consider the equations $\cale_\mu$. By the consideration above, it has the
form $h_x(h_x^2-h(x)^2+A_2)=2A_+\cosh(x)$. This equation has (at most) three
local branches of solutions satisfying the initial value problem $h|_{x=0}=0$
each corresponding to a root of the equation $\cale_\mu|_{x=0}$ considered as a
cubic polynomial on $h_x|_{x=0}$. However, we can immediately see that these
solutions are $2a^{(i)}\sinh(x)$ where $a^{(i)}$ are three roots of the
polynomial equation $a{\cdot}(4a^2+A_2)=A_+$.\footnote%
{\ The case when this polynomial has a multiple root is degenerate: In this
 case the discriminant of $\cale_\mu$ with respect to $h_x$ vanishes identically
 along the corresponding solution $h(x)$. Moreover, solving the corresponding
 initial value problem in the form of series $h(x)=\sum_jc_jx^j$ with $c_0=0$
 we obtain the cubic equation on $c_1{\cdot}(4c_1^2+A_2)=A_+$ which has one double
 and one simple root. Substituting this double root in $c_1$, all successive
 equations on $c_2,c_3,\ldots$ can be solved uniquely. This gives us the uniqueness
of the problem also for this degenerate case.}
The uniqueness of solutions of the initial value problem for ODEs
implies the equality $h(x)=2a^{(i)}\sinh(x)$. However, this contradicts to the
non-constancy of $R=h_{xxx}h_x-h_{xx}^2$. So finally we have excluded the
possibility $\lambda=±3\mu$ with $B_+\neq0\neq B_-$.

Let us notice that transforming the above solution $h(x)=2a\sinh(x)$ by means
of affine change of the coordinate $x$ ($x\mapsto b(x+x_0)$ with $b\neq0,x_0\in\cc$) and
the function itself ($h\mapsto a'h+c$ with $a'\neq0,c\in\cc$) we can obtain all
solutions given in items (1--3) of Theorem \ref{R=const}.

\medskip %
The next case we consider is when $B_-=0$. Recall that we also have $\mu=1$ and
$\lambda=3$. We shall consider the asymptotic behaviour of various expressions for
$x$ varying some on a real line in $\cc$ given by $\Im(x)=c$ and tending to
$\cplxi{\cdot}c-\infty$. To simplify notation, we write this as $x\to-\infty$.  Assume that
$A_-\neq0$.  Then from \eqref{eq-int} we obtain the asymptotic growth
$h(x)\sim-\big(\frac{3A_-}{C_0}\big)^{1/3}e^{-x/3}$ for $x\to-\infty$ and from
\eqref{eq-diff} a similar growth of the derivative $h_x$. The substitution of
this asymptotic in the l.h.s.\ of the equation $\cale_\mu$ would give the growth
$\sim C{\cdot}e^{-x}$ for $x\to-\infty$, whereas the l.h.s.\ $\cale_\mu$ decreases. The
contradiction shows that we must have $A_-=0$. Now from \eqref{eq-int} we
conclude that for $x\to-\infty$ the function $h(x)$ is given by the converging series
$h(x)=\sum_{j=0}^\infty a_je^{jx}$ with some complex coefficients $a_j$. The
substitution of this series in the equations gives
$a_1(B_2-9a_0^2)e^x+O(e^{2x})=0$ for $\cale_\lambda$, and
$(A_2a_1-a_0^2a_1-A_+)e^x+O(e^{2x})=0$ for $\cale_\mu$. In the case $a_1=0$ we
would have $A_+=0$ and hence $\Lambda_\lambda\equiv0$, which was excluded above.  Consequently,
$B_2=9a_0^2$. Substituting this relation in $\cale_\lambda$, we consider the further
expansion of $\cale_\lambda$. This gives us $a_1a_0^2e^{2x}+O(e^{3x})=0$, and hence
$a_0=0$ since by the above argument $a_1\neq0$. Repeating the substitution we
obtain $-(B_++8a_1^3)e^{3x}+O(e^{4x})=0$ from $\cale_\lambda$, and
$(A_2a_1-A_+)e^x+O(e^{2x})=0$ for $\cale_\mu$. This gives us $B_+=-8a_1^3$ and
$A_+=a_1A_2$. Now we see that for each of three roots $a_1$ of the equation
$8a_1^3+B_+=$ the function $a_1e^x$ satisfies the ODE $\cale_\mu$ and has the
correct asymptotic behaviour for $x\to-\infty$. Consequently, $h(x)$ is one of these
three solutions, and hence $R=h_{xxx}h_x-h_{xx}^2$ must be constant. The
obtained contradiction excludes the possibility $B_-=0$. 

\medskip %
Above we have proven the equality $\lambda=\mu$ under hypotheses of the theorem and
additional assumption $A_0\neq0\neq B_0$. Now we consider the case when one of these
coefficients vanishes, say $B_0=0$. Then $B_1\neq0$ since otherwise we obtain the
equation $B_0h_x=\Lambda_\lambda$ whose solutions are $h(x)=c_+e^{\mu x}+c_-e^{-\mu x}+c_0$
(or $h(x)=c_2x^2+c_1x+c_0$ in the case $\mu=0$) for which
$R=h_{xxx}h_x-h_{xx}^2$ would be constant. Making transformations $h\mapsto ah+c$
and $x\mapsto x/\lambda$ in the case $\lambda\neq0$ we change the equation $\cale_\lambda$ into
$h(x)h_x=B_+e^{2x}+B_-e^{-2x}$ (which means that we make $\lambda=2$) or
respectively $h(x)h_x=B_3x+B_4$. The integration gives
$h(x)^2=B_+e^{2x}-B_-e^{-2x}+B_5$ or resp.\ $h(x)^2=B_3x^2+2B_4x+B_5$
with some $B_5\in\cc$. As it was shown above without loss of
generality we can suppose that $B_+\neq0$. Hence we conclude that in the case
$\lambda=2$ the function $h(x)$ is given by a series
$h(x)=ae^x+\sum_{j=0}^\infty a_je^{-(2j+1)x}$ which converges for $\Re(x)\geq x_0\gg0$ and
such that $a^2=B_+$.  In the case $\lambda=0$ we obtain respectively
$h(x)=±\sqrt{B_3x^2+2B_4x+B_5}\sim±\sqrt{B_3}\,x$
in the case $\lambda=0$.  In particular, the asymptotic for the derivative is
$h_x\sim B_+^{1/2}e^{x}$ or respectively $h_x\sim±\sqrt{B_3}$.
The substitution in $\cale_\mu$ and comparing of the growth of the left and
right hand sides exclude the case $\lambda=0$ and shows that we could have $\mu=±1$,
$\mu=±2$, or $\mu=±3$ in the case $\lambda=2$. The case $\mu=±2=±\lambda$ is our claim, so we
must exclude two other possibilities.

First, we consider the case $\lambda=2$ and $\mu=±1$.
As above we distinguish the subcases $B_-\neq0$ and $B_-=0$ and start with the first
one $B_-\neq0$. Shifting the coordinate $x$ appropriately we make
$B_-=-B_+$. Then the coefficients $B_+,B_5$ and $a,a_0,a_1,\ldots$ are related as
\begin{equation}\label{koeffs}
\textstyle
B_+=a^2\quad B_5=-2aa_0\quad a_1=\frac{a^2-a_0^2}{2a}\quad 
a_2=-\frac{a_0(a^2-a_0^2)}{2a^2}\quad 
a_3=-\frac{(a^2-a_0^2)(a^2-5a_0^2)}{8a^3},
\end{equation}
and so on.  Substitute the series
$h(x)=ae^x+\sum_{j=0}^\infty a_je^{-(2j+1)x}$ and the above relation in $\cale_\mu$
and write the condition of the vanishing of the resulting expansion. We obtain
subsequently $A_+=aA_2-4a^2a_0$, $A_-=4aa_0^2-8a^2a_1-A_2a_0
=8aa_0^2-A_2a_0-4a^3$, %
and then the condition $\frac{3(a^2-a_0^2)(A_2-12aa_0)}{2a}=0$. So here we
have two possibilities: either $A_2=12aa_0$ or $a_0=± a$ (or both). However,
after substitution of the first relation $A_2=12aa_0$ in $\cale_\mu$ the first
non-trivial term will be $\frac{10(a^2-a_0^2)^2}{a}e^{-5x}$ which leads to the
relation $a_0=± a$ dropped above. But then all higher coefficients $a_1,a_2,\ldots$
must vanish and the solution $h(x)$ of the equation
$h(x)^2=B_+e^{2x}-B_-e^{-2x}+B_5$ will be $2a\cosh(2x)$ in the case $a_0=+a$
or respectively $2a\sinh(2x)$ in the case $a_0=-a$. In any case
$R=h_{xxx}h_x-h_{xx}^2$ will be constant. The obtained contradiction excludes
the possibility $\lambda=±2\mu$, $B_0=0$, and $B_-\neq0$.

Our next subcase is $\lambda=2$, $\mu=1$, and $B_0=B_-=0$. The procedure here is
essentially the same as in the previous subcase: Substituting the series
$h(x)=ae^x+\sum_{j=0}^\infty a_je^{-(2j+1)x}$ in $h(x)^2=B_+e^{2x}+B_5$ we obtain
the relations 
\begin{equation}\label{koeffs0}
\textstyle
B_+=a^2\quad B_5=-2aa_0 \quad
a_1=-\frac{a_0^2}{2a} \quad a_2=\frac{a_0^3}{2a^2}
\quad a_3=-\frac{5a_0^4}{8a^3},
\end{equation}
and so on. Next we substitute the series
$h(x)=ae^x+\sum_{j=0}^\infty a_je^{-(2j+1)x}$ and the obtained relations in
$\cale_\mu$ and get $A_+=aA_2-4a^2a_0$, $A_-=8aa_0^2-A_2a_0$, and then the
condition $\frac{3a_0^2(A_2-12aa_0)}{2a}=0$. As before, setting $A_2=12aa_0$
in $\cale_\mu$ we then obtain $\frac{10a_0^4}{a}=0$ which gives us the condition
$a_0=0$ dropped before. So we must have $a_0=0$ and $h(x)=ae^{x}$, and hence
$R=h_{xxx}h_x-h_{xx}^2$ will vanish identically.  The contradiction excludes
also this subcase.

\smallskip %
Next we consider the case case $\lambda=2$ and $\mu=±3$ and start with the subcase one
$B_-\neq0$. As in the case $\mu=±1$ above we can additionally assume $B_-=-B_+$.
Then we obtain the same expansion $h(x)=ae^x+\sum_{j=0}^\infty a_je^{-(2j+1)x}$ with
the same relations \eqref{koeffs}. Substituting them in $\cale_\mu$ we obtain
subsequently the relations $A_+=- 8a^3$, $A_1=0$, $A_2=12aa_0$,
$A_-=12a^2a_0-4a_0^3$ and then the condition
$\frac{15(a^2-a_0^2)^2}{a}=0$. As above, in both cases $a_0=± a$  all higher
coefficients $a_1,a_2,a_3,\ldots$ vanish, the solution $h(x)$ must be either
$2a\cosh(x)$ or $2a\sinh(x)$, and the function
$R=h_{xxx}h_x-h_{xx}^2$ will be constant.

In the subcase $\lambda=2$ and $\mu=±3$ and $B_-=0$ we obtain respectively first the
relations \eqref{koeffs0}, then subsequently the relations $A_+=- 8a^3$,
$A_1=0$, $A_2=12aa_0$, $A_-=-4a_0^3$, and then the condition
$\frac{15a_0^4}{a}=0$. The rest follows as in the case $\lambda=2$, $\mu=1$, and
$B_0=B_-=0$ considered above.

\medskip %
This finishes the proof of the fact that under the hypotheses of the theorem
one has the relation $\mu=±\lambda$. Now we show the complete assertion, namely, the
uniqueness of the equation up to constant factor. As before, we suppose that
$h(x)$ satisfies two equations, for which we maintain the above notation
$\cale_\mu,\cale_\lambda, \Lambda_\mu,\Lambda_\lambda,A_0,\ldots,B_0,\ldots,A_±,B_±$. Besides, we may assume the
equality $\mu=\lambda$.

Since $\mu=\lambda$, a linear combination of $\cale_\mu$ and $\cale_\lambda$ is again an
equation of the same form with the same $\mu$. In particular, we can replace
$\cale_\mu$ or $\cale_\lambda$ by such a linear combination. Consequently, we can
assume that $B_0=0$, and in the case $A_0=B_0=0$ we may also suppose that
$B_1=0$. However, in the latter case we would have
$B_2h_x=B_+e^{\mu x}+B_-e^{-\mu x}$ (resp.\ $B_2h_x=B_3x+B_4$ in the case $\mu=0$) and
hence $R=h_{xxx}h_x-h_{xx}^2$ would be constant. The contradiction shows that
we must have $A_0\neq0\neq B_1$. Normalising, we can make $A_0=1=B_1$.

First, let us consider  the case $\mu=\lambda\neq0$. Here we apply essentially the same arguments
as in the above cases $\lambda=2,\mu=1$ and $\lambda=2,\mu=3$. As we have shown above, making
appropriate transformations the equation $\cale_\lambda$ can be brought to the form
$h(x)h_x=B_+e^{2x}+B_-e^{-2x})$ with $B_+\neq0$, in particular, we make
$\mu=\lambda=2$. In this way we obtain the algebraic equation
$h(x)^2=B_+e^{2x}-B_-e^{-2x}+B_5$ and the asymptotic growth
$h(x)=ae^x+O(e^{-x})$ and $h_x=ae^x+O(e^{-x})$ for $x\to+\infty$ with $a^2=B_+\neq0$.
The substitution gives the growth $-3a^3e^{3x}+O(e^{2x})$ of the l.h.s.\ of
$\cale_\mu$, which contradicts to $\Lambda_\mu=A_+e^{2x}+A_-e^{-2x}$.

The argumentation in the case $\mu=\lambda=0$ is as follows. The equation
$h(x)h_x=B_3x+B_4$ integrates to $h(x)^2=B_3x^2+2B_4x+B_5$. An appropriate
affine transformation of $x$ and a rescaling of $h$ bring this equation into
one of the following forms: $h(x)^2=x^2+1$, $h(x)^2=x^2$, $h(x)^2=x$, or
$h(x)^2=1$. In the cases $h(x)^2=x^2$ and $h(x)^2=1$ the expression
$R=h_{xxx}h_x-h_{xx}^2$ vanishes in contradiction to the hypothesis of the
theorem. In the remaining cases the function $h(x)$ can not satisfy the
equation $h_x(h_x^2-A_1h(x)+A_2)=A_3x+A_4$. 

\medskip %
It remains to consider the case $A_0=0=B_0$ (``Darboux-superintegrable
case'').  Then both $A_1$ and $B_1$ must be non-zero since otherwise
$R=h_{xxx}h_x-h_{xx}^2$ would be constant as we have shown
above. Normalisation of the equations transforms them into
$h_x(h(x)+A_2)=A_+e^{\mu x}+A_-e^{-\mu x}$ (or $=A_3x+A_4$ in the case $\mu=0$) and
respectively $h_x(h(x)+B_2)=B_+e^{\lambda x}+A_-e^{\lambda x}$. The subsequent integration
gives 
\begin{equation}\label{A0=0int}
\begin{split}
\textstyle
\frac{h(x)^2}{2}+A_2h(x)&=
\textstyle
\frac{A_+}{\mu}e^{\mu x}-\frac{A_-}{\mu}e^{-\mu x}+A_5,\\
\textstyle
\frac{h(x)^2}{2}+A_2h(x)&=
\textstyle
\frac{A_3}{2}x^2+A_4x+A_5 
\qquad\text{in the case $\mu=0$,}\\
\textstyle
\frac{h(x)^2}{2}+B_2h(x)&=
\textstyle
\frac{B_+}{\lambda}e^{\lambda x}-\frac{B_-}{\lambda}e^{-\lambda x}+B_5.
\end{split}
\end{equation}
In the case $A_+=A_-=0$ the function $h(x)$ must be constant which contradict the
hypotheses of the theorem. So one of these coefficient must be non-zero, and
changing the sign of $\mu$ if needed we can suppose that $A_+\neq0$. By the same
argument $B_+$ is non-zero. Observe that the equations \ref{A0=0int} establish
an algebraic dependence between the functions $e^{\mu x}$ and $e^{\lambda x}$ in the
case $\mu\neq0\neq\lambda$, and between the functions $x$ and $e^{\lambda x}$ in the
case $\mu=0\neq\lambda$. This can be possible only if $\mu=±\lambda$. In this situation the
difference of the integrated equations \ref{A0=0int} is 
\[
\textstyle
(A_2-B_2)h(x)=\frac{A_+-B_+}{\mu}e^{\mu x}-\frac{A_--B_-}{\mu}e^{-\mu x}+(A_5-B_5)
\]
or respectively 
\[
\textstyle
(A_2-B_2)h(x)=\frac{A_3-B_3}{2}x^2+(A_4-B_4)x+(A_5-B_5)
\]
in the case $\lambda=\mu=0$. Now it is obvious that the triviality of these relations
is the only possibility to avoid the contradiction with the condition
$R=h_{xxx}h_x-h_{xx}^2\neq\const$. This means the desired proportionality of the
equations.

\smallskip
The theorem is proved.
\qed

\subsection{Real solutions.} \label{real-sols} Recall that the Principal
equations \eqref{Eqh1r}, \eqref{Eqh1mu0} have the following meaning: If a
surface metric $g$ admits a linear and a non-trivial cubic integral then it
has the form $h_x^{-2}(dx^2+dy^2)$ with a function $h(x)$ satisfying one of
these two equations with some \emph{complex} parameters $\mu,A_0,\ldots,A_4$. Of
course, we are interested only in solutions for which $h_x$ is real. In this
case $h(x)=h_1(x)+\cplxi{\cdot}c$ with some \emph{real} function $h_1(x)$ and a {
 real} constant $c$. Substituting  we see that $h_1(x)$ satisfies the
same equation with new parameters $A_0,\ldots,A_4$. Thus we can consider only real
solutions $h(x)$. 

\begin{thm}\label{real-h}  Assume that the equation \eqref{Eqh1all}
 with some \emph{complex} parameter $\mu$ and \emph{complex} coefficients admits
 a real-valued solution $h(x)$ such that $R=h_{xxx}{\cdot}h_x-h_{xx}^2$ is
 non-constant. Then $\mu$ is real or purely imaginary (or zero) and the equation
 is complex proportional to another equation \eqref{Eqh1all} with { the
  same parameter $\mu$ and with} \emph{real} coefficients $A_0,\ldots,A_4$.
\end{thm}

\proof The result follows immediately from Theorem \ref{thm-uniq} applied to
the equation \eqref{Eqh1all} and its complex conjugate.\qed

\section{ Number of independent cubic integrals. \\
Proof of Kruglikov's ``big gap'' conjecture. \\ 
Summary of the proof of the main theorem.}  \label{sec:5} 

\subsection{ Number of cubic integrals and  Kruglikov's ``big  gap'' conjecture.}
\label{Krug-1}

{ In \cite{Kr} Kruglikov conjectured that the dimension of the space of cubic
integrals of a surface metric $g$ of non-constant curvature is at most $4$.
In this section we prove this result for metrics satisfying the hypotheses of
the main theorem. Our proof applies also for Darboux-superintegrable
metrics, however, the result in the case is not new.}

\begin{thm}\label{thm-4} Let a function $h(x)$ satisfy one of the equations
 \eqref{Eqh1all} with complex parameters $\mu,A_0,\ldots,A_4$. Assume that
 $R:=h_{xxx}h_x-h_{xx}^2$ is non-constant. Set $H:=\frac12h_x^2(p_x^2+p_y^2)$.
 Then the space of complex-valued functions $F(x,y;p_x,p_y)$ that  are cubic
 in momenta $(p_x,p_y)$ and satisfy the equation $\{H,F\}=0$ is $4$-dimensional (as vector space over $\mathbb{C}$).
\end{thm}

\proof We distinguish two main cases: $\mu\neq0$ and $\mu=0$ and start with the first
one. Set $L:=p_y$. Then $\{H,L\}=0$. This gives us the following $4$ linearly
independent solutions of the equation $\{H,F\}=0$: $L^3,H{\cdot}L$, and
2-dimensional space of solutions $F$ given by the formulas \eqref{a0-a3new}.%
\footnote{The fact that the parameters $\mu,A_0,\ldots,A_4$ in Theorem \ref{main.th}
 are real plays no role here.}  So the theorem claims that there are no more
linearly independent solutions.

We call functions $F(x,y;p_x,p_y)$ satisfying the hypotheses of the theorem
(complex) cubic integrals (of the Hamiltonian $H$ given by the function
$h$). Denote by $\calf_h$ the space of complex cubic integrals. It was shown
by Kruglikov \cite{Kr} that the space $\calf_h$ is finite-dimensional.%
\footnote{The proof in \cite{Kr} is given for the case
 $H(x,y;p_x,p_y)=(dx^2+dy^2)/\lambda(x,y)$ with \emph{real} $\lambda(x,y)$. It works in
 our situation without changes.}  %
The Jacobi identity implies that the formula $\call:F\mapsto\{L,F\}$ induces a well
defined homomorphism $\call:\calf_h\to\calf_h$, see \S \ref{1.2}. Consider the decomposition of
$\calf_h$ into generalised eigenspaces of $\call$ and the corresponding Jordan
blocks. Then $L^3$ and $H{\cdot}L$ are eigenvectors with eigenvalue $0$. Further,
the functions $F_+$ and respectively $F_-$ given by formula \eqref{a0-a3new}
with $C_-=0$ and respectively $C_+=0$ are eigenvectors of $\call$ with
eigenvalues $±\mu$.

It follows immediately from Theorem \ref{thm-uniq} that the space $\calf_h$
contains no eigenvectors of $\call$ with eigenvalue $\lambda\neq±\mu$. Thus in the case
$\mu\neq0$ the assertion of the theorem is equivalent to the non-existence of a
generalised eigenvector of $\call$ with eigenvalue $±\mu$ and the Jordan block
$\smpmatr{ ±\mu&1\\0&±\mu}$. 

Assume the contrary.  Then there would exist cubic integrals $F_0,F_1\in\calf_h$
satisfying $\{L,F_1\}=±\mu F_1+F_0$ and $\{L,F_0\}=±\mu F_0$. Inverting the $y$-axis
we can change the sign. So we assume that we have $+\mu$ in the formulas. Recall
that $L=p_y$ correspond to the vector field $\frac{\partial}{\partial y}$.  Integrating the
equations above we obtain $F_0=e^{\mu y}G_0$ and $F_1=ye^{\mu y}G_0+e^{\mu y}G_1$
where $G_0,G_1$ are some complex functions of $(x,p_x,p_y)$ cubic in momenta
$(p_x,p_y)$ and independent of $y$. Since $F_0$ is a cubic integral and an
eigenvector of $\call$ with eigenvalue $\mu$, it has the form \eqref{a0-a3new}.
Since $\{F_0,H\}=\{H,e^{\mu y}G_0\}=0$, the equation $\{F_1,H\}=0$ now reads $\{e^{\mu
 y}G_1,H\}+\{y,H\}e^{\mu y}G_0=0$. Write $G_1=\sum_{j=0}^3b_j(x)p_x^{3-j}p_y^j$. Since
$\{y,H\}=-p_yh_x^2$, we obtain the equation
\begin{equation}\label{EqG1}
\{e^{\mu y}G_1,H\}-p_yh_x^2e^{\mu y}G_0=0
\end{equation}

Solving this equation we apply the same procedure as in §\,\ref{Case1}. The
bracket $\{e^{\mu y}G_1,H\}$ is given by \eqref{braFH} in which we need to replace
$a_j(x)$ by $b_j(x)$. Thus the equation \eqref{EqG1} is equivalent to 5
equations which are \emph{inhomogeneous} versions of 5 equations in
\eqref{braFH} with the r.h.s.-s given by $p_yh_x^2e^{\mu y}G_0$. As in
§\,\ref{Case1} we solve successively the first 3 of them and resolve $b_3(x)$ from
the last one. This gives the following formulas (compare with \eqref{a0-a3}):
\begin{equation}\label{b0-b3}
\textstyle 
\begin{split}
b_0(x) &\textstyle 
= B_0h_x^3\\
b_1(x) &\textstyle 
= (-(\mu B_0+A_0){\cdot}h(x)+\frac{B_1}{2\mu}){\cdot}h_x^2\\
b_2(x) &\textstyle 
=  \half{\cdot}(-(B_1+\frac{A_1}{\mu}){\cdot}h(x)
+ (\mu^2B_0+2\mu A_0){\cdot}h(x)^2
+ 3B_0h_x^2+ B_2){\cdot}h_x  \\
b_3(x) &\textstyle 
= \frac{1}{2\mu^2}{\cdot}(3{\cdot}h_x^2{\cdot}(\mu B_0-A_0)
-B_1{\cdot}h(x)+\mu^2{\cdot}(\mu B_0+A_0){\cdot}h(x)^2+(\mu B_2-A_2)){\cdot}h_{xx},
\end{split}
\end{equation}
Substituting them in the remaining term of \eqref{EqG1} we obtain the equation 
\begin{equation}\label{Eqh3B}
\begin{split} 
&(3{\cdot}(\mu B_0-A_0){\cdot}h_x^2   +\mu^2{\cdot}(\mu B_0+A_0){\cdot}h(x)^2
-\mu{\cdot}B_1{\cdot}h(x)+\mu{\cdot}B_2-A_2){\cdot}h_{xxx}\\
+\;&6{\cdot}(\mu B_0-A_0){\cdot}h_x{\cdot}h_{xx}^2
\ \ +\ (6{\cdot}\mu^2{\cdot}(A_0+\mu{\cdot}B_0){\cdot}h(x)-3{\cdot}\mu{\cdot}B_1){\cdot}h_x{\cdot}h_{xx}\\
+\;&63{\cdot}\mu^2{\cdot}(A_0+\mu{\cdot}B_0){\cdot}h_x^3 \\
+\;&(\mu^4{\cdot}(\mu{\cdot}B_0+3{\cdot}A_0){\cdot}h(x)^2-\mu^2{\cdot}(\mu{\cdot}B_1+2{\cdot}A_1){\cdot}h(x)+\mu^2{\cdot}(\mu{\cdot}B_2+A_2)){\cdot}h_x
\ \ =0
\end{split}
\end{equation}
which is the counterpart of \eqref{Eqh3}. As the equation \eqref{Eqh3}, the
above equation can be partially integrated in the sense that it can be written
in the form (compare with \eqref{Eqh-diff})
\begin{equation}\label{Eqh-diffB}
\begin{split}
\mu{\cdot}&\textstyle
\bigl(\frac{d^2}{dx^2} + \mu^2\bigr)
\bigl( h_x{\cdot}(h_x^2{\cdot}B_0+(\mu^2{\cdot}B_0 +2\mu{\cdot}A_0){\cdot}h(x)^2
-(B_1+\frac{A_1}{\mu}){\cdot}h(x)+B_2) \bigr)\\
+\;&\textstyle
\bigl(\frac{d^2}{dx^2} - \mu^2\bigl)
\bigr( h_x{\cdot}(h_x^2{\cdot}A_0+\mu^2{\cdot}A_0{\cdot}h(x)^2-A_1{\cdot}h(x)+A_2) \bigr)=0.
\end{split}
\end{equation}
Let us now observe that the other equation $\{F_0,H\}=\{H,e^{\mu y}G_0\}=0$ is
equivalent to the equation %
$\bigr( h_x{\cdot}(h_x^2{\cdot}A_0+\mu^2{\cdot}A_0{\cdot}h(x)^2- A_1{\cdot}h(x)+A_2)\bigr)=
A_3\frac{\sin(\mu x)}{\mu}+ A_4\,\cos(\mu x)$ %
with some constants $A_3,A_4$, and that the l.h.s.\ of this equation appears
in \eqref{Eqh-diffB}. Then 
\[\textstyle
\bigl(\frac{d^2}{dx^2}-\mu^2\bigl)
  \bigl(A'_3\,\sin(\mu x)+A_4\,\cos(\mu x)\bigr)=
 -2\mu^2\bigl(A'_3\sin(\mu x)+A_4\,\cos(\mu x)\bigr)
\]
(here we set $A'_3=\frac{A_3}{\mu}$) and so \eqref{Eqh-diffB} is equivalent to
\begin{equation}\label{Eqh1rB}
\begin{split}
\textstyle
h_x{\cdot}\bigl(h_x^2{\cdot}B_0+(\mu^2{\cdot}B_0 +2\mu{\cdot}A_0){\cdot}h(x)^2
-(B_1+\frac{A_1}{\mu}){\cdot}h(x)+B_2  \bigr)=&\\
(B_3+A_4{\cdot}x){\cdot}\sin(\mu x)-(A'_3{\cdot}x+B_4){\cdot}\cos(\mu x)&
\end{split}   
\end{equation}
For convenience in future let us make the substitution $\mu\mapsto\cplxi\mu$,
$B_i\mapsto\cplxi B_i$ in the equations \eqref{Eqh1rB} and \eqref{Eqh1r}, and rearrange their
r.h.s.-s. Then the equations transform into
\begin{align}
\label{Eqh1imA}
\textstyle
h_x{\cdot}\bigl(h_x^2{\cdot}A_0-\mu^2{\cdot}A_0{\cdot}h(x)^2
-A_1{\cdot}h(x)+A_2  \bigr)=
A_+{\cdot}e^{\mu x}+A_-{\cdot}e^{-\mu x}&\\
\begin{split}\label{Eqh1imB}
\textstyle
h_x{\cdot}\bigl(h_x^2{\cdot}B_0-(\mu^2{\cdot}B_0 -2\mu{\cdot}A_0){\cdot}h(x)^2
-(B_1-\frac{A_1}{\mu}){\cdot}h(x)+B_2  \bigr)=&\\
(B_++A_+{\cdot}x){\cdot}e^{\mu x}+(B_--A_-x){\cdot}e^{-\mu x}&
\end{split}   
\end{align}
Notice that the condition of non-triviality of $F_0$ is equivalent to the
non-vanishing of at least one parameter $A_0,A_1,A_2$.  Further, by Lemma
\ref{rhs=0} both $A_+$ and $A_-$ can not vanish together. Replacing $\mu$ by
$-\mu$, if needed, we can suppose that $A_+\neq0$.

We consider several subcases. The first one is $A_0=0$. Then the equation
\eqref{Eqh1imA} can be integrated as
\begin{equation}\label{Eqh1imA+-}
\textstyle
-\frac{A_1}{2}h(x)^2+A_2h(x)= 
\frac{A_+}{\mu}{\cdot}e^{\mu x}-\frac{A_-}{\mu}{\cdot}e^{-\mu x}+A_5
\end{equation}
If, moreover, $A_1=0$, then $h(x)= \frac{A_+}{\mu A_2}{\cdot}e^{\mu x}-\frac{A_-}{\mu
 A_2}{\cdot}e^{-\mu x}+\frac{A_5}{A_2}$, %
and then $R=\const$ in this case in contradiction with the hypotheses of the
theorem. Otherwise we make the substitution $x\mapsto2x/\mu$. After the substitution
$\mu$ transforms into $2$ and the r.h.s.\ of \eqref{Eqh1imA+-} into
$\frac{A_+}{2}{\cdot}e^{2x}-\frac{A_-}{2}{\cdot}e^{-2x}+A_5$. So we can conclude that
for $x\to+\infty$ the function $h(x)$ is given by the converging series
$ae^{x}+\sum_{j=0}^\infty a_je^{-jx}$. But then the substitution of this series in
\ref{Eqh1imB} gives the following leading terms for $x\to+\infty$: $-3a^3e^{3x}$
for the l.h.s., and $\frac{2A_+}{\mu}xe^{2\mu}$ for the r.h.s. The obtained
contradiction shows that the case $A_0=0$ is impossible.

In the case  $A_0\neq0$ we make the substitution $x\mapsto3x/\mu$ which makes $\mu=3$, and
subtract \eqref{Eqh1imA} from \eqref{Eqh1imB} with coefficients $B_0/A_0$.
This gives us the equation 
\begin{equation}\label{Eqh1C}
h_x(3C_0h(x)^2+2C_1h(x)+C_2)=A_+xe^{3x}+C_4e^{3x}+(C_5x+C_6)e^{-3x}
\end{equation}
with some constants $C_0,C_1,\ldots$ such that $C_0\neq0$.  Integrating it, we obtain
\begin{equation}\label{Eqh1Cint}
\textstyle
C_0h(x)^3+C_1h(x)^2+C_2h(x)+C_3=
\frac{A_+}{3}xe^{3x}+\wt C_4e^{3x}+(\wt C_5x+\wt C_6)e^{-3x}
\end{equation}
with some new constants $C_3,\wt C_5,\wt C_6$. From this equation we conclude
that $h(x)$ is a $3$-sheeted function on $\cc$ with a ramification on some
discrete subset $S\subset\cc$ and that for $x\to+\infty$ every branch of $h(x)$ has a
behaviour $h(x)=ax^{1/3}e^x(1+\Lambda(x))$ for some function $\Lambda(x)$ admitting a
converging series $\sum_{ij=0}^\infty a_{ij}e^{-ix}x^{-j/3}$ with
$a_{00}=0$. Moreover, the derivative $h_x$ is given by the derivative of the
expansion above and is a similar series $h_x= ax^{1/3}e^x(1+\wt\Lambda(x))$ with
$\wt\Lambda(x)= \sum_{ij=0}^\infty\ti a_{ij}e^{-ix}x^{-j/3}$ such that $\ti
a_{00}=0$. Substituting these expansions in 
\eqref{Eqh1imA} we obtain the term $-8a^3xe^{3x}$ for the l.h.s., which
contradicts to the growth $A_+e^{3x}$ of r.h.s. 

This prohibits the possibility $A_0\neq0$ for a solution $h(x)$ of the pair of
equations \eqref{Eqh1imA}--\eqref{Eqh1imB}, and thus excludes cubic integrals
$F_1,F_0$ such that $\{L,F_1\}=F_0$ and $\{L,F_0\}=0$.

\medskip %
Now we consider the case $\mu=0$. As above, denote $L=p_x$, set
$\call(F):=\{L,F\}$ for any function $F(x,y;p_x,p_y)$ and let $\calf_h$ be the
space of complex cubic integrals oh $H$. Then as above $\calf_h$ is finite
dimensional and $\call:\calf_h\to\calf_h$ is a well-defined homomorphism. It
follows from Theorem \ref{thm-uniq} that in the case $\mu=0$ the homomorphism
$\call:\calf_h\to\calf_h$ has unique eigenvalue $\mu=0$ and $\calf_h$ is a sum of
Jordan blocks with eigenvalue $\mu=0$. 

Notice that $L^3$ and $L{\cdot}H$ are eigenvectors of $\call$ with eigenvalue
$\mu=0$. We are going to prove that $\calf_h$ contains  only two linearly
independent eigenvectors, a unique Jordan block of
size $3{×}3$, and no other  Jordan blocks.

Let $F_0=F_0(x,y;p_x,p_y)$ be given by \eqref{a01-mu0} with coefficients
$a_0(x),\ldots,a_3(x)$ given by \eqref{a0-a3mu0old} with $\wt A_1=\wt A_3=0$.  Then
$F_0$ is a cubic integral, $\call(F_0)=A_1{\cdot}L^3+A_3{\cdot}L{\cdot}H\neq0$, and
$\call^2(F_0)=0$.  Assume that we have some other non-zero cubic integral $F'$
such that $\call^2(F')=0$. Then the calculation made in §\,\ref{1.2} and
§\,\ref{Case2} show that $F_0$ must be given by the same \eqref{a01-mu0} with
new coefficients $a'_0(x),\ldots,a'_3(x)$ given by \eqref{a0-a3mu0old} with parameters
$A'_0,\ldots,A'_3$ instead of $A_0,\ldots,A_3$, such that $h(x)$ satisfies the Principle
equation \eqref{Eqh1all} (iii) with $A_i$ replaced by $A'_i$.

At this point we obtain two subcases. The first is when both $A'_1,A'_3$
vanish, which means that $\call(F')=0$, i.e., $F'$ is an eigenvector of $\call$
with eigenvalue $\mu=0$. In this situation from the equation \eqref{Eqh1all}
(iii) we see that either $h_x$ is constant or all parameters $A'_0,\ldots,A'_4$
must vanish. The first possibility would yield $R=h_{xxx}h_x-h_{xx}^2\equiv0$,
which contradicts the hypotheses of the theorem. Thus all parameters
$A'_0,\ldots,A'_4$ must vanish, and then $F'=\wt A'_1{\cdot}L^3+\wt A'_3{\cdot}L{\cdot}H$, a
linear combination of $L^3$ and $L{\cdot}H$.

Let us underline, that the latter argument demonstrates that the space of
eigenvectors of $\call$ in $\calf_h$ is 2-dimensional with a basis $L^3,L{\cdot}H$.

The remaining subcase is when not all parameters $A'_0,\ldots,A'_4$ vanish and we
obtain a new equation of the form \eqref{Eqh1all} (iii). In this situation the
uniqueness from Theorem \ref{thm-uniq} ensures that $A'_i=c{\cdot}A_i$ with some
coefficient $c$. But in this case $F'=c{\cdot}F_0 + \wt A'_1{\cdot}L^3+\wt
A'_3{\cdot}L{\cdot}H$ %
with the same coefficient $c$ and some parameters $\wt A'_1,\wt A'_3$. This
demonstrates that the space of cubic integrals $F'$ satisfying
$\call^2(F')=0$ is 3-dimensional with a basis $F_0,L^3,L{\cdot}H$. In particular,
we can not have two distinct Jordan blocks. 

\smallskip %
Finally, let us show that there does exist a Jordan block of size $3{×}3$, and
no Jordan block of size $4{×}4$. For this purpose we try to find a cubic
integral $F$ satisfying $\call^4(F)=0$. Since the operator $\call$ acts as the
derivation in $y$, the condition  $\call^4(F)=0$ means that $F$ is a
polynomial in $y$ of degree $\leq3$.  This means
that we can write $F$ in the form
\begin{equation}\label{Fy3-mu0}
\textstyle
F=\sum_{i=0}^2\sum_{j=0}^3a_{ij}(x)y^ip_x^{3-j}p_y^j
\end{equation}
with some coefficients $a_{ij}(x)$. Writing down the equation $\{F,H\}=0$ and
considering its coefficients at monomials $y^ip_x^{4-j}p_y^j$ we obtain 15
ODEs on functions $a_{ij}(x)$ and $h(x)$. We solve them subsequently using the
conditions $h_x\neq0$, $h_{xx}\neq0$ and substituting the results in successive
equations. Doing so we obtain,%
\footnote{In calculation the authors used
 Maple\textsuperscript{®} software.}  the following formulas, in
which $a_{i,j;\,x}$ denote the derivatives of $a_{i,j}(x)$ and $A_{ij}$ are
integration constants:
\begin{enumerate}
\item \label{a00} $a_{0,0}(x) = A_{00}{\cdot}h_x^3$ \ from \
 $2{\cdot}h_x{\cdot}(-h_x{\cdot}a_{0,0;\,x}+3{\cdot}a_{0,0}(x){\cdot}h_{xx})=0$;
\item \label{a10} $a_{1,0}(x) = A_{10}{\cdot}h_x^3$ \ from \
 $2{\cdot}h_x{\cdot}(-h_x{\cdot}a_{1,0;\,x}+3{\cdot}h_{xx}{\cdot}a_{1,0}(x))=0$;
\item \label{a20} $a_{2,0}(x) = A_{20}{\cdot}h_x^3$ \ from \
 $2{\cdot}h_x{\cdot}(-h_x{\cdot}a_{2,0;\,x}+3{\cdot}h_{xx}{\cdot}a_{2,0}(x))=0$;
\item \label{a30} $a_{3,0}(x) = A_{30}{\cdot}h_x^3$ \ from \
 $2{\cdot}h_x{\cdot}(-h_x{\cdot}a_{3,0;\,x}+3{\cdot}h_{xx}{\cdot}a_{3,0}(x))=0$;
\item \label{a31} $a_{3,1}(x) = A_{31}{\cdot}h_x^2$ \ from \
 $2{\cdot}h_x{\cdot}(-h_x{\cdot}a_{3,1;\,x}+2{\cdot}a_{3,1}(x){\cdot}h_{xx})=0$;
\item \label{a13} $a_{1,3}(x) = \frac{h_{xx}}{h_x}{\cdot}a_{0,2}(x)$ \ from \
 $2{\cdot}h_x{\cdot}(-h_x{\cdot}a_{1,3}(x)+h_{xx}{\cdot}a_{0,2}(x))=0$;
\item \label{a23} $a_{2,3}(x) = \frac{h_{xx}}{2h_x}{\cdot}a_{1,2}(x)$ \ from \
 $2{\cdot}h_x{\cdot}(-2{\cdot}h_x{\cdot}a_{2,3}(x)+h_{xx}{\cdot}a_{1,2}(x))=0$;
\item \label{a33} $a_{3,3}(x) = \frac{h_{xx}}{3h_x}{\cdot}a_{2,2}(x)$ \ from \
 $2{\cdot}h_x{\cdot}(-3{\cdot}h_x{\cdot}a_{3,3}(x)+h_{xx}{\cdot}a_{2,2}(x))=0$;
\item \label{a32} $a_{3,2}(x) = 0$ \ from \ $2{\cdot}h_x{\cdot}h_{xx}{\cdot}a_{3,2}(x)=0$;
\item \label{a01} $a_{0,1}(x) = (-A_{10}{\cdot}h(x)+A_{01}){\cdot}h_x^2$ \ from \
 $2{\cdot}h_x{\cdot}(-h_x^4{\cdot}A_{10}-h_x{\cdot}a_{0,1;\,x}+2{\cdot}a_{0,1}(x){\cdot}h_{xx})=0$;
\item \label{a11} $a_{1,1}(x) = (-2{\cdot}A_{20}{\cdot}h(x)+A_{11}){\cdot}h_x^2$ \ from \
 $2{\cdot}h_x{\cdot}(-2{\cdot}h_x^4{\cdot}A_{20}-h_x{\cdot}a_{1,1;\,x}+2{\cdot}a_{1,1}(x){\cdot}h_{xx})=0$;
\item \label{a21} $a_{2,1}(x) = (-3{\cdot}A_{30}{\cdot}h(x)+A_{21}){\cdot}h_x^2$ \ from \
 $2{\cdot}h_x{\cdot}(-3{\cdot}h_x^4{\cdot}A_{30}-h_x{\cdot}a_{2,1;\,x}+2{\cdot}a_{2,1}(x){\cdot}h_{xx})=0$;
\item \label{a22} $a_{2,2}(x) =
 (-3A_{31}{\cdot}h(x)+\frac32A_{20}{\cdot}h_x^2+A_{22}){\cdot}h_x$ \ from \
 $2{\cdot}h_x{\cdot}(-3{\cdot}h_x^3{\cdot}A_{31}+3{\cdot}h_x^3{\cdot}h_{xx}{\cdot}A_{20}-h_x{\cdot}a_{2,2;\,x}
 +h_{xx}{\cdot}a_{2,2}(x))=0$;
\item[(*)] \label{A30} $A_{30}=0$ \ from \ $6{\cdot}h_x^4{\cdot}h_{xx}{\cdot}A_{30}=0$ at
 $y^3p_x^2p_y^2$; \ \
\item \label{a02} $a_{0,2}(x) =
 (A_{20}{\cdot}h(x)^2-A_{11}{\cdot}h(x)+\frac32A_{00}{\cdot}h_x^2+A_{02}){\cdot}h_x$ \ from \
 $2{\cdot}h_x{\cdot}(2{\cdot}h_x^3{\cdot}A_{20}{\cdot}h(x)-h_x^3{\cdot}A_{11}+3{\cdot}h_x^3{\cdot}h_{xx}{\cdot}A_{00}
 -h_x{\cdot}a_{0,2;\,x}+h_{xx}{\cdot}a_{0,2}(x))=0$;
\item \label{a12} $a_{1,2}(x) =
 (\frac{3}{2}A_{10}{\cdot}h_x^2-2A_{21}{\cdot}h(x)+A_{12}){\cdot}h_x$ \ from \
 $2{\cdot}h_x{\cdot}(2{\cdot}h_x^3{\cdot}A_{21}-3{\cdot}h_x^3{\cdot}h_{xx}{\cdot}A_{10}+h_x{\cdot}a_{1,2;\,x}
 -h_{xx}{\cdot}a_{1,2}(x))=0$.
\suspend{enumerate} The latter 4 relations yield certain correction in some
formulas above:
\resume{enumerate}
\item[(\ref{a21}')] $a_{2,1}(x) = A_{21}{\cdot}h_x^2 $;
\item[(\ref{a30}')] $a_{3,0}(x)=0$; 
\item[(\ref{a13}')] $a_{1,3}(x)=(A_{20}{\cdot}h(x)^2-A_{11}{\cdot}h(x)
+\frac32A_{00}{\cdot}h_x^2+A_{02}){\cdot}h_{xx}$; 
\item[(\ref{a23}')] $a_{2,3}(x)= 
(\frac{3}{4}A_{10}{\cdot}h_x^2 -A_{21}{\cdot}h(x)+\frac12A_{12}){\cdot}h_{xx}$; 
\item[(\ref{a33}')] $a_{3,3}(x)=  
(\frac12 A_{20}{\cdot}h_x^2-A_{31}{\cdot}h(x)+\frac13 A_{22}){\cdot}h_{xx}$.
\suspend{enumerate}
Finally, the following formula will be obtained later, we write it here simply
for completeness:
\resume{enumerate}
\item $a_{0,3}(x)=( -A_{10}{\cdot}h(x)+A_{01} ){\cdot}h_x^2
-\frac14A_{23}{\cdot}x^2 -\frac12A_{24}{\cdot}x  +A_{03}$.
\label{a03} 
\end{enumerate}

After calculation of (\ref{a00})--(\ref{a12}) and (*) it remains $4$
equations. One of them --- the coefficient at $y^0p_x^{\ }p_y^3$ --- can be
written as
\begin{equation}\label{Eqa03B}
\textstyle
 a_{0,3;x}= 2{\cdot}(A_{01}-A_{10}{\cdot}h(x)){\cdot}h_x{\cdot}h_{xx}
-\frac{3}{2}{\cdot}A_{10}{\cdot}h_x^3+(2A_{21}{\cdot}h(x)-A_{12}){\cdot}h_x
\end{equation}
and will be treated later. Three other are the coefficients at %
$y^1p_x^{\ }p_y^3$, $y^2p_x^{\ }p_y^3$ and $y^3p_x^{\ }p_y^3$. After
normalisation we obtain
\begin{align} 
\label{Eqh3B1}
\begin{split}
&\big( 3A_{00}{\cdot}h_x^2+ 2A_{20}{\cdot}h(x)^2
   -2A_{11}{\cdot}h(x)+2A_{02} \big){\cdot}h_{xxx}
+6A_{00}{\cdot}h_{xx}^2{\cdot}h_x\\
&\ \ +6{\cdot}\big(2A_{20}{\cdot}h(x)-A_{11} \big){\cdot}h_x{\cdot}h_{xx}
+6A_{20}{\cdot}h_x^3 +4{\cdot}\big(A_{22}-3A_{31}{\cdot}h(x)\big){\cdot}h_x=0
\end{split}
\\\label{Eqh3B2}
&\big(3A_{10}{\cdot}h_x^2-4A_{21}{\cdot}h(x)+2A_{12}\big){\cdot}h_{xxx}
+6A_{10}{\cdot}h_{xx}^2{\cdot}h_x-12A_{21}{\cdot}h_x{\cdot}h_{xx}=0
\\\label{Eqh3B3}
&\big(3A_{20}{\cdot}h_x^2-6A_{31}{\cdot}h(x)+2A_{22}\big){\cdot}h_{xxx}
+6A_{20}{\cdot}h_{xx}^2{\cdot}h_x-18A_{31}{\cdot}h_x{\cdot}h_{xx}=0
\end{align}
Double integration of the latter two gives
\begin{align} 
\label{Eqh1B2}
h_x{\cdot}(A_{10}{\cdot}h_x^2-4A_{21}{\cdot}h(x)+2A_{12})= A_{23}x+A_{24},
\\
\label{Eqh1B3}
h_x{\cdot}(A_{20}{\cdot}h_x^2-6A_{31}{\cdot}h(x)+2A_{22})=A_{33}x+A_{34},
\end{align} 
where $A_{23},A_{24},A_{33},A_{34}$ are new integration constants.
Substituting the latter relation in the equation \eqref{Eqa03B} we can
integrate it getting the formula $(\ref{a03})$ for $a_{0,3}(x)$ promised
above. The equation \eqref{Eqh3B1} can not be integrated. However, we can do
this even twice with the linear combination
``\eqref{Eqh3B1}''$-2{\cdot}$``\eqref{Eqh1B3}'', and the result is the equation
\begin{equation}
\label{Eqh1B1}
\textstyle
h_x{\cdot} (A_{00}{\cdot}h_x^2+2A_{20}{\cdot}h(x)^2 -2A_{11}{\cdot}h(x)+2A_{02})
=-\frac{1}{3}A_{33}{\cdot}x^3-A_{34}{\cdot}x^2+A_{13}{\cdot}x+A_{14}
\end{equation}
with new integration constants $A_{13},A_{14}$. 

\smallskip %
From the above calculation we conclude the following: The metric
$g=h_x^{-2}(dx^2+dy^2)$ admits a cubic integral of the form \eqref{Fy3-mu0} if
and only if the $h(x)$ satisfies the equations \eqref{Eqh1B1}, \eqref{Eqh1B2}, and
\eqref{Eqh1B3}, and then the integral can be constructed using the formulas
(\ref{a00})--(\ref{a03}) for the coefficients $a_{i,j}(x)$. 

\smallskip %
Observe that the non-zero coefficients $a_{3,j}(x)$ are
$a_{3,1}(x)=A_{31}h_x^2$ and $a_{3,3}(x)=(\frac12
A_{20}{\cdot}h_x^2-A_{31}{\cdot}h(x)+\frac13 A_{22}){\cdot}h_{xx}$. This means that the
the non-existence of a Jordan block of size $4{×}4$ is equivalent to the
vanishing of all three coefficients $A_{20},A_{22},A_{31}$. Suppose for a
moment that this is not the case. Then the equation \eqref{Eqh1B3} is
non-trivial. Since we are looking for functions $h(x)$ for which
$R=h_{xxx}h_x-h_{xx}^2$ is non-constant, Lemma \ref{rhs=0} says that
either  $A_{13}$ or $A_{14}$ (or both) is non-zero. Consequently, the equation
\eqref{Eqh1B1} is also non-trivial. 

We distinguish several possibilities. The first is when $A_{20}=0$. Then
integrating \eqref{Eqh1B3}, we obtain
\begin{equation} 
\label{Eqh1B3int}
\textstyle
-3A_{31}{\cdot}h(x)^2+2A_{22}{\cdot}h(x)=\frac12 A_{33}x^2+ A_{34}x+A_{35}
\end{equation}
with an integration constant $A_{35}$.  We can not have $A_{31}=0$ since in
this case $R=h_{xxx}h_x-h_{xx}^2$ would be constant. Also we can not have
$A_{33}=A_{34}=0$ by the same reason. In the remaining cases we study the
behaviour of (branches of) the function $h(x)$ considered as a solution of an
\emph{algebraic} equation \eqref{Eqh1B3int}. It follows that for $x\to\infty$ the
solution $h(x)$ grows like $\sim ax+O(x^0)$ (case $A_{33}\neq0$) or like
$\sim ax^{1/2}+O(x^0)$ (case $A_{33}=0$).  From \eqref{Eqh1B3int} we obtain the
growth of $h_x$: $\sim a+O(x\inv)$ in the case $A_{33}\neq0$ and
$\sim ax^{-1/2}+O(x\inv)$ in case $A_{33}=0$. In any of these two cases we see
that the growth of the l.h.s.\ of \eqref{Eqh1B1} is slower than that of the
r.h.s. The contradiction shows that we must have $A_{20}\neq0$.

Now subtract the equation \eqref{Eqh1B3} from \eqref{Eqh1B1} with coefficient 
$\frac{A_{00}}{A_{20}}$. We obtain the equation
\begin{equation}
\label{Eqh1B1-B3}
\textstyle
h_x{\cdot} (2A_{20}{\cdot}h(x)^2 +A'_{11}{\cdot}h(x)+A'_{02})
=-\frac{1}{3}A_{33}{\cdot}x^3-A_{34}{\cdot}x^2+A'_{13}{\cdot}x+A'_{14}
\end{equation}
with new constants $A'_{11},A'_{02},A'_{13},A'_{14}$ and the same constants
$A_{20}\neq0,A_{33},A_{34}$. Integration gives
\begin{equation}
\label{Eqh1B1-B3int}
\textstyle
\frac{2A_{20}}{3}{\cdot}h(x)^3 +\frac{A'_{11}}{2}{\cdot}h(x)^2+A'_{02}{\cdot}h(x)
=-\frac{A_{33}}{12}{\cdot}x^4-\frac{A_{34}}{3}{\cdot}x^3
+\frac{A'_{13}}{2}{\cdot}x^2+A'_{14}x+A_{15}
\end{equation}
with an integration constant $A_{15}$. Our next possibility is $A_{33}\neq0$. In
this case the same argument as above gives the asymptotic
$h(x)\sim ax^{4/3}+O(x)$ and $h_x\sim\frac{4a}{3}x^{1/3}+O(x^{0})$ for
$x\to\infty$. Moreover, the coefficient $a$ satisfies the relation
$A_{33}=-8A_{20}{\cdot}a^3$. Substituting these asymptotic in \eqref{Eqh1B3}
gives the asymptotic $-8A_{31}{\cdot}a^2{\cdot}x^{5/3}+O(x^{4/3})$ for the l.h.s.\
provided $A_{31}\neq0$. This is a contradictions. Consequently, we must have
$A_{31}=0$. But then we obtain the asymptotic
$\frac{64}{27}A_{31}{\cdot}a^3{\cdot}x+O(x^{2/3})$ for the l.h.s., whereas on the
r.h.s.\ we have the leading term $A_{33}x=-8A_{20}{\cdot}a^3x$. As the result we
conclude that the case $A_{33}\neq0$ is impossible and we must have $A_{33}=0$.

Notice, that in this remaining case $A_{33}=0$ we must have $A_{34}$ since
otherwise $R=h_{xxx}h_x-h_{xx}^2$ would be constant by Lemma \ref{rhs=0}. Here
as above from \eqref{Eqh1B1-B3int} we can conclude the asymptotic
$h(x)=ax+O(x^0)$ and $h_x=a+O(x\inv)$ for $x\to\infty$. The substitution of these
asymptotic in \eqref{Eqh1B3} gives linear growth $-6A_{31}{\cdot}a{\cdot}x+O(x^0)$
for the l.h.s.\ provided $A_{31}\neq0$, whereas the r.h.s.\ is constantly
$A_{34}$. So we must have $A_{31}=0$. But the only solutions of the equation
\eqref{Eqh1B3} with $A_{31}=A_{33}=0$ are $h_x=\const$ which contradict the
hypothesis $R=h_{xxx}h_x-h_{xx}^2\neq\const$. The latter contradiction
demonstrates the non-existence of Jordan blocks of size $4{×}4$.

\medskip %
Finally, we describe cubic integrals $F$ which give Jordan blocks of size
$3{×}3$. The latter condition means that $F$ is given by \eqref{Fy3-mu0} with
vanishing coefficients $a_{3,j}(x)$, $j=0,\ldots,3$. From formulas
(\ref{a00})--(\ref{a03}) we see that this condition is equivalent to vanishing
of parameters $A_{20},A_{31},A_{22}$.  In this case we see from \eqref{Eqh1B3}
that the coefficients $A_{13},A_{14}$ must also vanish. Thus the equation
\eqref{Eqh1B3} becomes trivial, and the equation \eqref{Eqh1B1} simplifies to
\begin{equation}
\label{Eqh1B1y2}
\textstyle
h_x{\cdot} (A_{00}{\cdot}h_x^2 -2A_{11}{\cdot}h(x)+2A_{02})= A_{13}{\cdot}x+A_{14}.
\end{equation}
The uniqueness of the equation proved in Theorem \ref{thm-uniq} ensures that
the equation \eqref{Eqh1B2} and \eqref{Eqh1B1y2} must be proportional to each
other and to the equation \eqref{Eqh1all} (iii). Let us denote by $C_1,C_2$
the corresponding proportionality coefficients. This leads to the following
relations:
\begin{equation}\label{Acsubs}
\begin{split} 
&\textstyle
A_{00} = C_1{\cdot} A_{0},\quad A_{11}=\half{\cdot}C_1{\cdot} A_{1},\quad 
A_{02}=\half{\cdot}C_1{\cdot} A_{2},\quad A_{13} =C_1{\cdot}A_{3},\quad 
A_{14} =C_1{\cdot}A_{4},\quad 
\\
&\textstyle
A_{10} = C_2{\cdot}A_{0},\quad 
A_{21}=\frac14{\cdot}C_2{\cdot} A_{1},\quad A_{12}=\half{\cdot}C_2{\cdot} A_{2},\quad 
A_{23}=C_2{\cdot}A_{3},\quad A_{24} =C_2{\cdot}A_{4}.
\end{split} 
\end{equation}
Additionally set $A_{03}=C_{L3}$ and $A_{01} = \half C_{LH}$. Substituting these
relations and formulas  (\ref{a00})--(\ref{a03}) we obtain the following
formula for a cubic integral $F$ in the case $\mu=0$:
\begin{align} \label{F-L3LH}
F\;& \textstyle
= C_{L3}{\cdot}p_y^3 +C_{LH}{\cdot}\half h_x^2(p_x^2p_y^{\,}+p_y^3) \\
 \label{Flong-C1}
\begin{split} 
 &\textstyle
+C_1{\cdot}\Big(  A_0{\cdot}h_x^3{\cdot}p_x^3
+\frac{y}{2}{\cdot}A_1{\cdot}h_x^2{\cdot}p_x^2p_y^{\,}
+\half{\cdot}(3A_0{\cdot}h_x^2-A_1{\cdot}h(x)+A_2){\cdot}h_x{\cdot}p_y^2{\cdot}p_x \\
 &\textstyle \qquad \qquad 
\frac{y}{2}{\cdot}(3A_0{\cdot}h_x^2-A_1{\cdot}h(x)+A_2){\cdot}h_{xx}{\cdot}p_y^3 
\Big)
\end{split}\\
 \label{Flong-C2}
\begin{split} 
 &\textstyle
+C_2{\cdot}\Big( A_0{\cdot}h_x^3{\cdot}y{\cdot}p_x^3
+(\frac{y^2}{4}{\cdot}A_1{\cdot}h_x^2-A_0{\cdot}h_x^2{\cdot}h(x)){\cdot}p_x^2p_y^{\,}
\\ &\textstyle \qquad 
+\half{\cdot}(3A_0{\cdot}h_x^2-A_1{\cdot}h(x)+A_2){\cdot}h_x{\cdot}y{\cdot}p_x^{\,}p_y^2\\
 &\textstyle
\qquad +\big(\frac14{\cdot}(3A_0{\cdot}h_x^2-A_1{\cdot}h(x)+A_2){\cdot}h_{xx}{\cdot}y^2
-A_0{\cdot}h_x^2{\cdot}h(x) -\half{\cdot}A_4{\cdot}x-\frac14{\cdot}A_3{\cdot}x^2\big){\cdot}p_y^3 
\Big).
\end{split}
\end{align}
Thus $F$ is a linear combination of $4$ independent cubic integrals as stated
in Theorem \ref{main.th}, two of which are $L^3$ and $L{\cdot}H$, and other two
are linear and quadratic in $y$. 

Notice that the formulas \eqref{Flong-C1} and \eqref{Flong-C2} for cubic
integrals differ from those in \eqref{Fmu0} and \eqref{a0-a3mu0}. To obtain
the last ones we need to subtract from \eqref{Flong-C1} and \eqref{Flong-C2}
the derivative of the equation \eqref{Eqh1all} (iii) with coefficients 
$\frac{y}{2}{\cdot}p_y^3$ and respectively $\frac{y^2}{4}{\cdot}p_y^3$.
\qed

\subsection{  {Summing up: Theorem \ref{main.th} is proved.}
}

As we explained in §§\ref{1.2}, \ref{Case1}, \ref{Case2}, any metric $g$
admitting linear integral $L$ and cubic integral $F$ such that $L$, $H$ and
$F$ are functionally independent have in an { appropriate} coordinate
system the form $h_x^{-2}(dx^2 + dy^2) $, where { the function $h(x)$}
satisfies \eqref{Eqh1r} or \eqref{Eqh1mu0}.  

In Theorem \ref{real-h} we proved that $\mu$ in \eqref{Eqh1r} is real, or pure
imaginary,  and the parameters $A_0,...,A_4$ in \eqref{Eqh1r}
and in \eqref{Eqh1mu0} { become real after multiplication with an
 appropriate constant.}
Thus, the equations (\ref{Eqh1r}, \ref{Eqh1mu0}) are essentially the equations
\eqref{Eqh1all}. 

It follows from §§\ref{1.2}, \ref{Case1}, \ref{Case2} that the metric
$g=h_x^{-2}(dx^2+dy^2)$ admits at least one cubic integral $F_1$ which has the
form from Theorem \ref{main.th}, and two functionally independent integrals
$F_1,F_2$ in the case $\mu\neq0$.  In Section \ref{sec:5} we construct an
additional independent cubic integral $F_2$ in the case $\mu=0$, and explained
why there are no other integrals (unless the metric has constant curvature).

\smallskip
Theorem \ref{main.th} is proved.

\section{Global solutions}%
\label{globalism} %
In this section we show  that if the function $h(x)$ satisfies the
equation \eqref{Eqh1all} (ii) and $h'(x_0)>0$ at some point $x_0$ whereas the
real parameters $\mu>0,A_0,\ldots,A_4$ satisfy inequalities $A_0>0$, $\mu{\cdot}A_4>|A_3|$
then the metric $g=\frac{1}{h_x^2}(dx^2+dy^2)$ smoothly extends to the sphere
$S^2$ together with the linear integral $L=p_y$ and the cubic integral $F$
given by \eqref{a0-a3-ell}.  More precisely, we show that if $(r,\varphi)$ are polar
coordinats on $\rr^2$ related to $(x,y)$ by $r=e^{x/\mu}, \varphi=y/\mu$, then $g$, $L$,
and $F$ are well-defined on the punctured plane $\rr^2\bs\{0\}$ and extend
smoothly to the origin $0$ and to the infinity point $\infty$ of the Riemann
compactification $S^2=\overline{\mathbb{C}}=\rr^2\cup\{\infty\}$, such that the extended tensor $g$ is still a
(non-degenerate) Riemannian metric on the whole sphere $S^2$.

The obtained in this way family of examples of superintegrable metrics on the
sphere is new. 
Indeed, by \cite{Ki}   Darboux-superintegrable metrics can not live on a closed manifold, so the only known superintegrable metrics on the 2-sphere are the standard metrics of constant curvature. In view of Corollary \ref{generic-sol}, for most values of the parameters satisfying the above conditions, the metrics are not metrics of constant curvature.

Let us  describe the conditions on parameters $\mu>0,A_0,\ldots,A_4$ which
distinguish our global solutions.  Since we want that the linear integral $L$
to be also globally defined on $S^2$, the Killing vector field must be as in the standard 
rotation (see for example \cite{Bo-Ma-Fo}),   and we must have the elliptic case , i.e., $h$ must satisfy
\eqref{Eqh1all}, (ii). Since  Darboux-superintegrable case is impossible on closed surfaces  due to \cite{Ki},
$A_0$ is non-zero. Then dividing the equation by $A_0$
we obtain $A_0=1$. Applying the action $h(x)\mapsto h(x)+c$ we can make
$A_1=0$. Since $\mu\neq0$, making an appropriate rescaling in $x$ we can make
$\mu=1$.  Further, we rewrite the free term $A_3\sinh(x)+A_4\cosh(x)$ in the
form $A_+e^x+A_-e^{-x}$ and we impose the positivity condition $A_+,A_->0$.
This means that the free term is positive for all $x\in\rr$. An appropriate
translations in $x$ direction transforms the term $A_+e^x+A_-e^{-x}$ into
$A_e{\cdot}(e^x+e^{-x})$.  It is easy to see that in terms of the original
parameters our conditions are
\begin{equation}\label{cond-A-mu}
\mu>0,\quad A_0>0,\quad A_4{\cdot}\mu- |A_3|>0.
\end{equation}

\begin{lem}\label{loc-sol}   For any $A_e>0$, $A_2$ and  $h(x_0)$  there
 exists a unique real-analytic \emph{local} 
 solution $h(x)$  of the ODE
\begin{equation}\label{Eqh-gl}
\cale:= h_x(h_x^2-h(x)^2+A_2)- A_e{\cdot}(e^x+e^{-x})=0
\end{equation}
with the initial value $h(x_0)$ such that $h_x(x_0)$ is the unique
\emph{positive} root of the characteristic polynomial
$\chi(\lambda):=\lambda(\lambda^2+A_2-h(x_0)^2)- A_e{\cdot}(e^{x_0}+e^{-x_0})$.
\end{lem} 

\proof Considering the graph of the polynomial $\lambda(\lambda^2+a)$ for different $a$
(mainly, for the cases $a\geq0$ and $a<0$) we see that there exists a unique
positive solution of the equation $\lambda(\lambda^2+a)=A$ with $A>0$ which depends
real-analytically on $a$ and $A>0$.  The standard theory of ODEs implies now
the desired local existence and uniqueness of solutions of
\eqref{Eqh-gl}. 
\qed

\smallskip%
\state Remark. Notice that depending on coefficients $A>0$ and $a$ one
 could have two distinct, one double, or none \emph{negative} roots of the
 equation $\lambda(\lambda^2+a)=A$.  Vice versa, in the case $A<0$ there is one negative
 root and there could be two distinct, one double, or none positive roots.
 This explains our ``Ansatz'': we need that the right hand side \ 
 $A_3\sinh(x)+A_4\cosh(x)=A_+e^x+A_-e^{-x}$ remains positive.

\medskip%
Let $h(x)$ be a local solution constructed in the previous lemma. Make the
substitution $x=\log(t)$. Then the equation \eqref{Eqh-gl} transforms into
\begin{equation}\label{Eqh-t}
\wt\cale:= t{\cdot}h_t\big( (t{\cdot}h_t)^2-h(t)^2+A_2)- A_e{\cdot}(t+t\inv)=0
\end{equation}
defined for all $t>0$.

\begin{prop}\label{poles}  
 Any local solution $h(t)$ of \eqref{Eqh-t} with $h_t$ positive extends to the
 interval $t\in(0,+\infty)$. Moreover, the functions $t{\cdot}h(t)$, $t^2h_t$, and
 $t^3h_{tt}$ extend real-analytically to some interval $t\in(-\varepsilon,\varepsilon)$, $\varepsilon>0$. In
 particular, the function $h(t)$ has a simple pole at $t=0$.
\end{prop}

Notice that in terms of the variable $x$ the first assertion of the
proposition states that every local  solution $h(x)$ as in Lemma \ref{loc-sol} is
global, i.e., is defined for all $x\in\rr$.

\proof Considering the local behaviour of $h(t)$ at $t=0$, we show a stronger
property, namely, that $h(t)=f(t^2)/t$ for some real-analytic function $f(\tau)$,
defined in some interval $\tau\in[-\varepsilon,+\infty]$.  This means that we make the
substitution $t=\sqrt{\tau}$, or equivalently, $t^2=\tau$ or
$x=\frac12\log{\tau}$. However, we consider all three functions $h(x)$, $h(t)$,
and $h(\tau)$ as a single object given in different coordinates.

For $A$ positive and $h,A_2$ real, denote by $\eta=\eta(h,A_2;A)$ the unique
positive root of the polynomial $\lambda(\lambda^2+A_2-h^2)-A$. Then $\eta(h,A_2;A)$ is
monotone in every its argument $A>0, A_2$ and $h\neq0$: $\eta(h,A_2;A)$ will
increase if we increase $A>0$ and the absolute value $|h|$ and decrease
$A_2$. Consequently, for any given $A_2$ and $A_e>0$ any solution $h(x)$ of
\eqref{Eqh-gl} with positive $h_x$ satisfies $h_x\geq\eta^* :=\eta(0,A_2,;2A_e)$. In
particular, $h(x)$ is monotone.

Now let us observe the following two facts. First, the translation in $x$
(which means the multiplication of $t$ by a constant) transforms our equation
\eqref{Eqh-gl}
into 
\begin{equation}\label{Eqh-gl-tr}
h_x(h_x^2-h(x)^2+A_2)- A_+{\cdot}e^x-A_-{\cdot}e^{-x}=0
\end{equation}
or respectively \eqref{Eqh-t} into 
\begin{equation}\label{Eqh-t-tr}
t{\cdot}h_t\big( (t{\cdot}h_t)^2-h(t)^2+A_2)
- A_+{\cdot}t- A_-{\cdot}t\inv=0
\end{equation}
with positive parameters $A_+,A_-$. Second, for any values of parameters our
equation has a very simple function as the solution. Namely, the function %
$\ti h(t):=C_h(-A_-/t+A_+{\cdot}t)$ satisfies the equation \eqref{Eqh-t-tr} if and
only if the constant $C_h$ is the unique positive root of the polynomial
$C_h(4A_+A_-{\cdot}C_h^2+A_2)=1$. We use these special solutions and their
translations in $x$ ($\Leftrightarrow$ reparametrisations $t\mapsto c{\cdot}t$) to estimate the
behaviour of a general solution $h(t)$.

Let $h(t)$ be any solution of \eqref{Eqh-t} defined in a neighbourhood of the
initial value $t_0$.  Denote $h_0:=h|_{t=t_0}$. Let $\ti h(t)$ be the unique
solution of \eqref{Eqh-t-tr} of the form $\ti h(t):=C_h(-A_-/t+A_+{\cdot}t)$ with
$C_h>0$. In the case when $\ti h(t_0)=h_0$ we must have $h(t)=\ti h(t)$
everywhere and thus $h(t)$ is defined globally on $(0,+\infty)$.

Assume that $\ti h(t_0)>h_0$. Since both functions are solution of the same
1st order ODE, $\ti h(t)>h(t)$ for every $t$ from the maximal existence
interval $(t_-,t_+)\subset(0,+\infty)$ of the solution $h(t)$. Further, since $h(t)$ is
monotone increasing and  bounded from above by the function $\ti h(t)$
defined on the whole ray $(0,+\infty)$, we conclude that the solution $h(t)$ does
not explode and exists on the whole interval $(t_0,+\infty)$. In particular, the
existence interval for $h(t)$ is $(t_-,+\infty)$. Moreover, $h(t)\leq C_hA_+t$ for all
$t\in(t_-,+\infty)$.

Our next step is construction of a similar lower bounding function $h_-(t)$
which also has the form $h_-(t)=C_h{\cdot}(-A'_-/t+A_+{\cdot}t)$ and satisfies the
equation \eqref{Eqh-t-tr} with some new parameters $A'_-$ and $A_2$. This
means that constructing $h_-(t)$ from $\ti h(t)$ we change only one parameter,
namely $A_-$. Increasing it the value $h_-(t_0)$ will decrease. Thus we can
find the new value $A'_->A_-$ from the condition $h_-(t_0)=h_0$. Then we find
the new value $A'_2$ from the relation $C_h(4A_+A'_-{\cdot}C_h^2+A'_2)=1$. Notice
that since $A'_-$ was increased, $A'_2$ is lower than $A_2$, i.e., $A'_2<A_2$.
Assume additionally that $h_0\leq0$. Then $h(t)<0$ for every $t\in(t_0,t_0)$. Now
using the monotonicity on the function $\eta(h,A_2;A)$ and comparing the
equations for $h(t)$ and $h_-(t)$ and the initial values $h_-(t_0)=h(t_0)$ and
$\frac{d}{dt}h_-(t_0)>\frac{d}{dt}h(t_0)$, we conclude the following: for
every $t$ in the whole existence interval $(t_-,t_0)$ left from the initial
point $t_0$ we have $h_-(t)<h(t)$, $|h_-(t)|^2>|h(t)|^2$, and
$\frac{d}{dt}h_-(t)>\frac{d}{dt}h(t)$.

The mononoticity argument above can be applied in the case when $h_0>0$. In
this case we need an additional step. Namely, since we have the uniform
estimate $h'(t)\geq\eta^*>0$, our solution $h(t)$ vanishes at the unique $t_1$ lying
in the existence interval $(t_-,t_0)$ left from $t_0$. Then we apply the same
argument above at the point $t_1$ instead of $t_0$.

Finally, recall that above we have proceeded under assumption %
$\ti h(t_0)>h_0$. The remaining case $\ti h(t_0)<h_0$ is treated quite
similarly, and we only indicates the changes. In this new situation the
function $\ti h(t):=C_h(-A_-/t+A_+{\cdot}t)$ will be a lower bound for our
solution $h(t)$, i.e., $h(t)>\ti h(t)$ everywhere in the existence interval.
This fact together with monotonicity of $h(t)$ will imply the extensibility of
$h(t)$ left until the value $t_-=0$. To construct the upper bounding function
$h^+(t)$ we increase the parameter $A_+$ and also decrease $A_2$. Moreover, in
the case $h_0<0$ we need an additional step, in which we go right to the
unique point $t_1\in(t_0,t_+)$ such that $h(t_1)=0$, and apply the argument at
this point $t_1$.

For the second assertion of the proposition in the case $\ti h(t_0)<h_0$ we need one
bounding function, namely, the function $h_+(t)$ of the familiar form
$h_+(t):=C_h(-A'_-/t+A_+{\cdot}t)$ such that $h_+(t)>h(t)$ for $t<t_0$ ($h^+(t)$
does this for $t>t_0$ or respectively for $t>t_1>t_0$). As before in the case
$h_0>0$ we go left to some point $t_2<t_0$ such that $h(t_2)<0$. Notice that
$\ti h(t_2)$ is still less than $h(t_2)$, and hence $\ti h(t_2)<0$. Now, if we
start to decrease $A_-$, the value $\ti h(t_2)$ will increase until it
arrives to $0$. Consequently, for some $A'_0\in(0,A_0)$ the function
$h_+(t):=C_h(-A'_-/t+A_+{\cdot}t)$ has value $h_+(t_2)=h(t_2)$. The same
monotonicity argument as above gives us $h(t)<h_+(t)$ for $t\in(0,t_2)$.

This gives the desired global existence of the solution $h$: the maximal
existence interval is $(0,+\infty)$ for the coordinate $t$ which means that $h(x)$
exists for all $x\in\rr$.

\smallskip %
As a consequence of the above argument, we obtain the estimate
$-\frac{C}{t}<h(t)<-\frac{c}{t}$ for $0<t\ll1$ with some positive constants
$c,C$ for any solution of \eqref{Eqh-t}.

Now let us make the substitution $t=\sqrt\tau$ and
$h(t)=f(t^2)/t=f(\tau)/\sqrt\tau$. Then the equation \eqref{Eqh-t} transforms into
\begin{equation}\label{Eqf-tau} 
8{\cdot}f_\tau^3{\cdot}\tau^2+(-12{\cdot}f_\tau^2{\cdot}f(\tau)
-A_e+2{\cdot}f_\tau{\cdot}A_2){\cdot}\tau-f(\tau){\cdot}A_2-A_e+4{\cdot}f(\tau)^2{\cdot}f_\tau
=0
\end{equation}
For $\tau$ small enough and $f<0$ we can resolve this equation with respect to
the derivative $f_\tau$, and then \eqref{Eqf-tau} transforms into the form
\begin{equation}\label{Eqf-tau-expl} 
f_\tau = \Phi(\tau,f(\tau);A_2,A_e)
\end{equation}
for some real-analytic function $\Phi(\tau,f;A_2,A_e)$ of arguments $\tau\in[-\varepsilon,\varepsilon]$,
$f<0$, $A_2,A_e$. In particular, $\Phi|_{\tau=0}=\frac{A_e+A_2{\cdot}f}{4f^2}$. It
follows, that for every $A_2,A_e$, every negative $f_0$, and every $\tau_0$ small
enough there exists the unique solution of the equation \eqref{Eqf-tau} with
the initial value $f(\tau_0)=f_0$. More precisely, for given intervals
$A_2,A_e\in[-C,C]$, $f_0\in[-C,-c]$ with $0<c<C$ there exists an $\varepsilon=\varepsilon(c,C)>0$ such
that the equation \eqref{Eqf-tau} has the unique solution with the initial
value $f(\tau_0)=f_0$ at $\tau_0\in[-\varepsilon,\varepsilon]$ such that
$f_\tau(\tau_0)=\Phi(\tau_0,f_0;A_2,A_e)$. More over, this solution is well-defined and
real-analytic on the whole interval $\tau\in[-\varepsilon,\varepsilon]$.

Next, we observe that the above estimate $-\frac{C}{t}<h(t)<-\frac{c}{t}$ for
$0<t\ll1$ is equivalent to the estimate $-C<f(\tau)<-c$ for $0<\tau\ll1$. The
proposition follows. %
\qed

\medskip%
Now we state the result about the extensibility of $g$ and $F$ to a
 metric and a cubic integral defined globally on $S^2$.

\begin{thm} \label{glob.th} %
For any $A_e>0,A_2$ and $h_0$, let $h(t)$ be the unique solution
 of the equation \eqref{Eqh-t} with the initial value $h|_{t=1}=h_0$ and with
 $h_t|_{t=1}>0$. Then the metric 
\begin{equation}\label{g-glob}
g= \frac{dt^2+t^2{\cdot}d\varphi^2}{t^4h_t^2}
\end{equation}
defined on the plane $\rr^2$ with the polar coordinates $(t,\varphi)$ extends to a
real analytic metric on the sphere $S^2=\overline{\mathbb{C}}=\rr^2\cup\{\infty\}$ with the Killing vector
$\mbfv=\frac{\partial}{\partial\varphi}$ which admits a cubic integral $F$ { also well-defined
 and real-analytic globally on $S^2$.}

Moreover, the metric $g$ has constant curvature if and only if $h_0=0$.
\end{thm}

\proof By Proposition \ref{poles} the function $h(t)$ is well-defined for all
$t\in(0,+\infty)$ and for $t$ small enough $h(t)=\frac{f(t^2)}{t}$ with some
real-analytic function $f(\tau)$ such that $f(0)<0$. Consequently, in a
neighbourhood of the origin the function $t^2h_t$ is real-analytic and
non-vanishing. It follows that the formula
$g=\frac{dt^2+t^2{\cdot}d\varphi^2}{t^4h_t^2}$ defines a non-degenerate real-analytic
Riemannian metric in a neighbourhood of the origin in $\rr^2$ with the polar
coordinates $(t,\varphi)$. Substitution $t=e^x$, $\varphi=y$ transforms this metric into
the familiar form $g=\frac{dx^2+dy^2}{h_x^2}$ and the equation \eqref{Eqh-t}
into \eqref{Eqh-gl}. By Proposition \ref{poles} the metric
$g=\frac{dx^2+dy^2}{h_x^2}$ is well-defined for all $x\in\rr$ or equivalently
for all $t\in(0,+\infty)$. This means that the metric
$g=\frac{dt^2+t^2{\cdot}d\varphi^2}{t^4h_t^2}$ is well-defined on the whole plane
$\rr^2$.

To show the extensibility to the infinity point $\infty$ we apply the inversion of
the sphere $S^2=\rr^2\cup\{\infty\}$ with respect to the unit circle given by the
condition $t=1$. Recall that the inversion map interchanges the origin $0$ and
the infinity point $\infty$ and that in the polar coordinates it is given by
$(t,\varphi)\mapsto(t\inv,\varphi)$. Changing to the coordinates $x=\log(t),y=\varphi$ we obtain the
formula $(x,y)\mapsto(-x,y)$. So we conclude immediately that the extensibility of
the metric $g$ to the infinity $\infty$ is equivalent to the extensibility to the
origin $0$, and thus this is the case.

Let $\xi:=t{\cdot}\cos(\varphi)$ and $\eta:=t{\cdot}\sin(\varphi)$ be the Cartesian coordinates
corresponding to the polar coordinates $(t,\varphi)$ and $p_\xi,p_\eta$ the corresponding
momenta, i.e., dual coordinates on $T^*S^2$.  Then the vector field $\partial_y=\partial_\varphi$
is given by $\xi\partial_\eta-\eta\partial_\xi$. This means that the linear integral $L=p_y$ is given
by $L=\xi p_\eta-\eta p_\xi$ and hence $L$ extends smoothly to the origin. By the symmetry
argument $L$ extends also to the infinity point $\infty$.

\smallskip %
It remains to show that the cubic integral $F$ given by \eqref{a0-a3-ell} also
extends to the origin and to the infinity point $\infty$. Our argumentation is as
follows. First, we substitute in the formulas \eqref{a0-a3-ell} our values of
parameters $A_0=\mu=1$ and $A_1=0$. Next, without loss of generality we may set
$\phi=0$ for ``phase parameter'' in \eqref{a0-a3-ell}.  Then each function
$a_0(x),\ldots,a_3(x)$ in \eqref{a0-a3-ell} becomes a sum of linear and cubic
monomials in functions $h(x),h_x,h_{xx}$. Now we observe that each function
$h(x),h_x,h_{xx}$, considered as a function of the variable $t$, has the form
$f_i(t^2)/t$ for certain real-analytic function $f_i(\tau)$, namely, 
$f_0(\tau)=f(\tau)$ as in the proof of Proposition \ref{poles},
$f_1(\tau)=2\tau f'(\tau)-f(\tau)$ and $f_2(\tau)=4\tau^2f''(\tau)+f(\tau)$. Further, in the same way
as it was done for $p_y=\xi\partial_\eta-\eta\partial_\xi$, we obtain the formula $p_x=\xi p_\xi+\eta p_\eta$.
Substituting all these relations and also the relations
$\cos(y)=\frac\xi t,\sin(y)=\frac\eta t$ in \eqref{a0-a3-ell} we see that in
coordinates $(\xi,\eta)$ the integral $F$ has the form
\[\textstyle
F(\xi,\eta;p_\xi,p_\eta)= \sum_{i=0}^3\sum_{j=0}^4 
p_\xi^{3-i}p_\eta^i\,\xi^{4-j}\eta^j\,\frac{\psi_{ij}(\tau)}{\tau^2}
\]
where $\tau=t^2=\xi^2+\eta^2$ and $\psi_{ij}(\tau)$ are some real-analytic functions.  It
follows that to establish the real-analyticity of $F$ we need to know only the
linear parts of the functions $\psi_{ij}(\tau)$. Considering the expressions of
these linear parts of $\psi_{ij}(\tau)$ in terms of the the functions
$h(x),h_x,h_{xx}$ we see that only two lower monomials of each function
$h(x),h_x,h_{xx}$ are involved. We write
$h(t)=-c_0{\cdot}t\inv+c_1{\cdot}t+O(t^3)$. Differentiation yields $h_x
=t{\cdot}h_t=+c_0{\cdot}t\inv+c_1{\cdot}t+O(t^3)$ and
$h_{xx}=t{\cdot}\frac{d}{dt}(t{\cdot}h_t)=-c_0{\cdot}t\inv+c_1{\cdot}t+O(t^3)$. Substituting
these expressions in the equation \eqref{Eqh-t} and considering the
coefficients by $t^{±1}$, we obtain algebraic equations
\[
4{\cdot}c_0^2{\cdot}c_1+c_0{\cdot}A_2-A_e=0, \qquad 
4{\cdot}c_1^2{\cdot}c_0+c_1{\cdot}A_2-A_e=0.
\]
So we conclude the equality $c_1=c_0$ and the formula $A_e = c_0{\cdot}(4{\cdot}c_0^2+A_2)$.
Now, substituting all these formulas in \eqref{a0-a3-ell} 
we obtain
\[\textstyle
\begin{split} 
F& \textstyle
= c_0^3{\cdot}p_\xi{\cdot}(p_\xi^2+p_\eta^2)
+(3{\cdot}c_0^3{\cdot}\xi^2+ \frac{c_0{\cdot}(A_2+6{\cdot}c_0^2)}{2} {\cdot}\eta^2){\cdot}p_\xi^3
-c_0{\cdot}(2{\cdot}c_0^2+A_2){\cdot}\eta{\cdot}\xi{\cdot}p_\xi^2{\cdot}p_\eta\\
& \textstyle
+( \frac{c_0{\cdot}(10{\cdot}c_0^2+A_2)}{2} {\cdot}\xi^2
+c_0^3{\cdot}\eta^2){\cdot}p_\eta^2{\cdot}p_\xi
+2{\cdot}c_0^3{\cdot}\eta{\cdot}\xi{\cdot}p_\eta^3
+O(\xi^4+\eta^4)
\end{split} 
\]
in which $O(\xi^4+\eta^4)$ means term of higher degree in $\xi,\eta$.

This shows that the cubic integral $F$ extends real-analytically to the origin
as desired. The extensibility of $F$ to the infinity point $\infty$ can be obtained
from the extensibility to the origin by means of the inversion.

The theorem follows.
\qed

\section{Conclusion}\label{conclu}

We found all two-dimensional Riemannian metrics whose geodesic flows admit one
integral linear in momenta $(L)$ and one integral cubic in momenta $(F)$ such
that $L,F$ and the Hamiltonian $H$ of the geodesic flow are functionally
independent.  Within these metrics, we point out the metrics that are already
known, and proved that most of our metrics are new. { We have also showed
 that,  in the case when the parameters satisfy certain inequalities,  the metric
 and the integrals $L$ and $F$ extend real-analytically to  the sphere $S^2$,
 giving  new unexpected  examples  of  integrable metrics on the sphere.}

The results and the methods of our paper suggest the following directions of
further investigations.

\begin{prob}  \label{p1}
Generalise our result for integrals of higher degree.  \end{prob}

In other words, we suggest to construct all two-dimensional metrics whose
geodesic flows admit one integral linear in momenta $(L)$ and one integral
polynomial in momenta of degree 4, (5, 6, etc.)  in momenta $(F)$ such that
$L, F$ and $H$ are functionally independent.

The main trick that allowed us to solve the case (linear integral $L$ + cubic
integral $F$) survives in this setup: the Poisson bracket $\{L, F\}$ is again an
integral of the same degree as $F$.  Arguing as in §\ref{1.2} one can reduce
the problem to analyse of certain system of ODE. Though it is not clear in
advance whether one can reduce this system of ODE to one equation (as we did
in the case (linear integral $L$ + cubic integral $F$)), the approach should
at least allow to construct new examples of superintegrable metrics.

\begin{prob} Generalize our results for pseudo-Riemannian metrics. \end{prob}

We expect that this is possible to do  the local description using the same idea.  We do not expect that one can find  the analog of our global examples on closed surfaces in the pseudo-Riemannian case. Generally, it could be complicated to generalize global Riemannian construction to the pseudo-Riemannian 
setting. In certain cases though the existence of  additional structure such as additional integrals (as for example in \cite{Ma3}) allows to keep control over the situation.

\begin{prob} Quantise the cubic integral.  \label{quantize}  \end{prob}

Take a metric $g$ from Theorem \ref{main.th} and consider its Laplacian $\Delta$
(since our metric is $\tfrac{1}{h_x^2}(dx^2 + dy^2)$, \ $\Delta= {h_x^2}
\left(\tfrac{\partial^2 }{\partial^2x} + \tfrac{\partial^2 }{\partial^2y} \right) $; one can view it as a
mapping $\Delta: C^{\infty}(M^2)\to C^{\infty}(M^2)$ though one also can consider Laplacian as
a linear operator on bigger function spaces).

 Does there exist  a differential operator $\tilde F$  of degree $3$, 
\begin{eqnarray*} 
\tilde  F &= & a_0(x,y) \tfrac{\partial^3 }{\partial^3x}  
+ a_1(x,y) \tfrac{\partial^2 }{\partial^2x}  \tfrac{\partial }{\partial y} + a_2(x,y) \tfrac{\partial }{\partial x}  \tfrac{\partial^2 }{\partial^2 y}  + a_3(x,y)   \tfrac{\partial^3 }{\partial^3 y} \cr  &+& b_0(x,y) \tfrac{\partial^2 }{\partial^2x}  +b_1(x,y) \tfrac{\partial  }{\partial x}  \tfrac{\partial }{\partial y} + b_2(x,y) \tfrac{\partial^2 }{\partial^2 y} + c_0(x,y) \tfrac{\partial }{\partial x}  +c_1(x,y)   \tfrac{\partial }{\partial y} + d(x,y)\end{eqnarray*} 
such that $\tilde F$ commute with $\Delta$, i.e., such that  for every smooth function $f:M^2\to \mathbb{R}$ 
$$
[\Delta, \tilde F](f):= \Delta(\tilde F(f))- \tilde F(\Delta(f))\equiv 0, $$
and such that its symbol 
$$ a_0(x,y) p_x^3   + a_1(x,y) p_x^2  p_y + a_2(x,y) p_x  p_y^2  + a_3(x,y)   p_y^3$$
coincides with the integral $F$ from Theorem \ref{main.th}? 

Recall that for all previously known superintegrable systems such quantisation
of the integral was possible, and was extremely useful for the describing of
eigenfunctions of $\Delta$. Note that  quantum superintegrability can be most effectively used (if it exists) in the case of metrics from Theorem \ref{glob.th}, since in this case the Laplacian is a selfadjoint operator. 

\begin{prob}\label{Realisation}
 Find   physical or mechanical systems realizing the Hamiltonian
  systems corresponding to (at least some) metrics constructed in
 Theorem \ref{main.th}.
\end{prob}
This problem is an interesting challenge for both mathematicians and  physicists,
especially in the case of global systems given by Theorem \ref{glob.th}.  Let us note that many 
classical examples of global integrable systems have arisen as the
mathematical models for concrete naturally defined dynamical systems
in physics and mechanics, and that many superintegrable metrics have physical realisation or 
can be applied  to solving of physical problems.

\begin{prob}   \label{besse} Describe the metrics from Theorem \ref{glob.th} in the terms of \cite[Chapter 4]{Be}. 
\end{prob} 

As we already mentioned above, all geodesics of the metrics from Theorem \ref{glob.th} are closed. 
By construction, the metrics are the metrics of revolution. Then, these metrics are a subclass of the so-called Tannery metrics from   \cite[Chapter 4]{Be}. 

\begin{prob}  Find isometric  imbeddings  of metrics from Theorem  \ref{glob.th} in 
$(\mathbb{R}^3, g_{\textrm{standard}})$. 
\end{prob}

Such isometric imbeddings  are  possible at least for certain metrics from \ref{glob.th}, since they have positive curvature as small perturbations of the standard metric. This problem is related 
 to   Problem \ref{Realisation}, and is a geometric analog  of it. It also is related to Problem \ref{besse} in view of  \cite[Chapter 4(C)]{Be}.

\subsection*{ Acknowledgements.}
We thank Ian Andreson, Boris Kruglikov, Ian Marquette, Manuel F.  Rañada, and Pavel
Winternitz for useful discussions and Deutsche Forschungsgemeinschaft
(priority program 1154 --- Global Differential Geometry and research training
group 1523 --- Quantum and Gravitational Fields) and FSU Jena for partial
financial support.


\

\ifx\undefined\bysame
\newcommand{\bysame}{\leavevmode\hbox to3em{\hrulefill}\,}
\fi

\def\entry#1#2#3#4\par{\bibitem[#1]{#1}
{\textsc{#2 }}{\sl{#3} }#4\par\vskip2pt}

\def\noentry#1#2#3#4\par{}

\def\mathrev#1{{{\bf Math.\ Rev.:\,}{#1}}}

\end{document}